\documentclass[aps,10pt,prc,twocolumn,tightenlines,preprintnumbers,showpacs,superscriptaddress,notitlepage,nofootinbib,floatfix,longbibliography]{revtex4-1}

\usepackage{amssymb}   \usepackage{amsmath}
\usepackage{dsfont}
\usepackage{CJKutf8}   \usepackage{graphicx}  \usepackage{dcolumn}   \usepackage{bm}        \usepackage{xspace}
\usepackage{standalone}
\usepackage{enumitem}
\usepackage{color}
\usepackage{xcolor}
\usepackage{slashed}
\usepackage{booktabs}
\usepackage{multirow}
\usepackage{stackrel}
\usepackage{rotating}
\usepackage{pifont}
\usepackage{mathtools}
\usepackage{simplewick}
\usepackage{float} \usepackage{tikz} \usepackage{hyperref}
\hypersetup{
    colorlinks=true,       linkcolor=blue,          citecolor=blue,        filecolor=blue,      urlcolor=blue           }
 \def\chroma{\texttt{Chroma}\xspace}

\def\eqref#1{{(\ref{#1})}}
\newcommand{\eqnref}[1]{Eq.~\eqref{#1}}

\newcommand{\figref}[1]{Fig.~\ref{#1}}

\newcommand{\secref}[1]{Sec.~\ref{#1}}
\newcommand{\appref}[1]{App.~\ref{#1}}
\newcommand{\tabref}[1]{Table~\ref{#1}}

\definecolor{kngrey}{HTML}{A6AAA9}
\definecolor{knred}{HTML}{EC5D57}
\definecolor{knorange}{HTML}{F39019}
\definecolor{knyellow}{HTML}{F5D328}
\definecolor{kngreen}{HTML}{70BF41}
\definecolor{knblue}{HTML}{51A7F9}
\definecolor{knpurple}{HTML}{B36AE2}

\def\ket#1{{|#1\rangle}}
\def\bra#1{{\langle #1|}}

\def\d{{\delta}}
\def\D{{\Delta}}

\def\g{{\gamma}}
\def\G{{\Gamma}}

\def\O{{\Omega}}
\def\S{{\Sigma}}
\def\s{{\sigma}}
\def\t{{\tau}}

\def\tsep{t_{\rm sep}}
\def\tinc{\tau_{\rm inc}}

\newcommand{\ithems}{
    Interdisciplinary Theoretical and Mathematical Sciences Program (iTHEMS),
    RIKEN, 2-1 Hirosawa,
    Wako, Saitama 351-0198, Japan
}
\newcommand{\qtlab}{
    Theoretical Quantum Physics Laboratory,
    RIKEN, 2-1 Hirosawa,
    Wako, Saitama 351-0198, Japan
}
\newcommand{\bochum}{
    Institut f{\"u}r Theoretische Physik II,
    Ruhr-Universit{\"a}t Bochum, D-44780 Bochum, Germany
}

\newcommand{\umich}{
	Physics Department,
	University of Michigan,
	Ann Arbor, MI 48109, USA
	}
\newcommand{\glasgow}{
 School of Physics and Astronomy,
    University of Glasgow,
    Glasgow G12 8QQ, UK
 }
\newcommand{\jlabt}{
	Theory Center,
	Thomas Jefferson National Accelerator Facility,
	Newport News, VA 23606, USA
	}

\newcommand{\lblnsd}{
    Nuclear Science Division,
    Lawrence Berkeley National Laboratory,
	Berkeley, CA 94720, USA
	}

\newcommand{\llnl}{
	Physics Division,
	Lawrence Livermore National Laboratory,
	Livermore, CA 94550, USA
	}
\newcommand{\llnldesign}{
	Design Physics Division,
	Lawrence Livermore National Laboratory,
	Livermore, CA 94550, USA
	}

\newcommand{\nvidia}{
    NVIDIA Corporation,
    2701 San Tomas Expressway, Santa Clara, CA 95050, USA
    }

\newcommand{\ucas}{
	Department of Physics,
	University of Chinese Academy of Sciences,
	Beijing, 100049, P. R. China
}
\newcommand{\ucb}{
	Department of Physics,
	University of California,
	Berkeley, CA 94720, USA
	}
\newcommand{\ucr}{
    Escuela de F\'isica , Universidad de Costa Rica,
    San Jos\'e, San Pedro, 11501, Costa Rica
}
\newcommand{\umd}{
	Department of Physics,
	University of Maryland,
	College Park, MD 20742, USA
}
\newcommand{\unc}{
	Department of Physics and Astronomy,
	University of North Carolina,
	Chapel Hill, NC 27516-3255, USA
	}
\newcommand{\uw}{
    Department of Physics,
    University of Washington,
    Seattle, WA 98195, USA
    }
\newcommand{\wm}{
	Department of Physics,
	The College of William \& Mary,
	Williamsburg, VA 23187, USA
	}
 \newcount\hour \newcount\hourminute \newcount\minute
\hour=\time \divide \hour by 60
\hourminute=\hour \multiply \hourminute by 60
\minute=\time \advance \minute by -\hourminute

\begin{document}

\preprint{JLAB-THY-21-3350}
\preprint{RIKEN-iTHEMS-Report-21}

\title{Detailed analysis of excited state systematics in a lattice QCD calculation of $g_A$}

\author{Jinchen~He (\begin{CJK*}{UTF8}{gbsn}何晋琛\end{CJK*})}
\affiliation{\ucas}
\affiliation{\ucb}

\author{David~A.~Brantley}
\affiliation{\llnl}

\author{Chia~Cheng~Chang (\begin{CJK*}{UTF8}{bsmi}張家丞\end{CJK*})}
\affiliation{\ithems}
\affiliation{\lblnsd}
\affiliation{\ucb}

\author{Ivan~Chernyshev}
\affiliation{\ucb}
\affiliation{\uw}

\author{Evan~Berkowitz}
\affiliation{\umd}

\author{Dean~Howarth}
\affiliation{\llnl}
\affiliation{\lblnsd}

\author{Christopher~K\"orber}
\affiliation{\bochum}
\affiliation{\lblnsd}

\author{Aaron~S.~Meyer}
\affiliation{\ucb}
\affiliation{\lblnsd}

\author{Henry~Monge-Camacho}
\affiliation{\unc}
\affiliation{\ucr}

\author{Enrico~Rinaldi}
\affiliation{\umich}
\affiliation{\qtlab}
\affiliation{\ithems}

\author{Chris Bouchard}
\affiliation{\glasgow}

\author{M.A.~Clark}
\affiliation{\nvidia}

\author{Arjun~Singh~Gambhir}
\affiliation{\llnldesign}
\affiliation{\lblnsd}

\author{Christopher~J.~Monahan}
\affiliation{\wm}
\affiliation{\jlabt}

\author{Amy~Nicholson}
\affiliation{\unc}
\affiliation{\lblnsd}

\author{Pavlos~Vranas}
\affiliation{\llnl}
\affiliation{\lblnsd}

\author{Andr\'{e}~Walker-Loud}
\affiliation{\lblnsd}
\affiliation{\llnl}
\affiliation{\ucb}

\collaboration{CalLat Collaboration}
\thanks{Not all in California}
\homepage{https://callat-qcd.github.io}

\date{\today}

\begin{abstract}
Excited state contamination remains one of the most challenging sources of systematic uncertainty to control in lattice QCD calculations of nucleon matrix elements and form factors:
early time separations are contaminated by excited states and late times suffer from an exponentially bad signal-to-noise problem.
High-statistics calculations at large time separations $\gtrsim1$ fm are commonly used to combat these issues.
In this work, focusing on $g_A$, we explore the alternative strategy of utilizing a large number of relatively low-statistics calculations at short to medium time separations (0.2--1 fm), combined with a multi-state analysis.
On an ensemble with a pion mass of approximately 310 MeV and a lattice spacing of approximately 0.09 fm, we find this provides a more robust and economical method of quantifying and controlling the excited state systematic uncertainty.
A quantitative separation of various types of excited states
enables the identification of the transition matrix elements as the dominant contamination.
The excited state contamination of the Feynman-Hellmann correlation function is found to reduce to the 1\% level at approximately 1 fm while for the more standard three-point functions,
this does not occur until after 2 fm.
Critical to our findings is the use of a global minimization, rather than fixing the spectrum from the two-point functions and using them as input to the three-point analysis.
We find that the ground state parameters determined in such a global analysis are stable against variations in the excited state model, the number of excited states, and the truncation of early-time or late-time numerical data.
\end{abstract}

\maketitle

\section{\label{sec:intro} Introduction}
Lattice QCD (LQCD) calculations of nucleon matrix elements have reached a level of maturity for inclusion in the most recent Flavour Lattice Averaging Group (FLAG) review~\cite{Aoki:2019cca}.
Results are now commonly obtained
%It is now common that results are obtained
with multiple lattice spacings, multiple volumes and pion masses at or near the physical pion mass.  Control over the continuum, infinite volume and physical pion mass extrapolations are necessary to compare LQCD results amongst themselves as well as with experiments.

However, there is an additional source of systematic uncertainty in the calculations which must be brought under control before the extrapolations can be relied upon, and that is the excited state contamination of the correlation functions.
The source of the issue is tied to the well-known signal-to-noise (S/N) problem~\cite{Lepage:1989hd}.
At early time, where the stochastic noise is under control, the correlation functions have significant contamination from excited states, while at large time, where there is ground state saturation, the noise overwhelms the signal.
To date, there are no calculations of nucleon three-point functions performed at light pion masses with sufficient statistics that the ground state matrix element can be determined at sufficiently large times that excited state contributions can be neglected -- the equivalent of a single exponential fit to the two-point function to extract the ground state energy.

The FLAG review summarizes challenges in controlling the excited state contamination in these calculations and the strategies various groups use to do so.
In this article, we will focus on two important points that are not discussed in the FLAG review and which are generally lacking in the literature:
\begin{enumerate}[leftmargin=*]
\item Stability of the ground state matrix element under the truncation of data and/or the variation in the number of excited states used in the analysis;

\item Quantification of the excited state contamination.
\end{enumerate}
The first issue is generally not discussed because most calculations utilize too few values of fixed source-sink separation times ($\tsep$) for data truncation to be feasible.
A larger number of matrix elements are required to perform multi-state fits, so a limited number of $\tsep$ values used limits the number of excited states that can be removed.
In the flavor physics LQCD community, stability of the extracted ground state observable over an appreciable time range (sufficiently large to rule out correlated fluctuations) has long been recognized as crucial for ruling out excited state contamination in two-point and three-point calculations~\cite{Lepage:2001ym, Wingate:2002fh, Gamiz:2009ku, Bazavov:2016nty, Bazavov:2018kjg}; any residual contamination to the ground state observable from excited states will either be observed as a trend in time of its value, or is smaller than the precision with which the observable has been extracted. For nucleon correlation functions, it is even more crucial that this stability be demonstrated as they are more susceptible to correlated late-time fluctuations through their more severely degrading S/N ratios. Care must be taken, however, to demonstrate this stability before the exponential growth of the noise erases the ability to detect time dependence in the observable.

The second issue is typically addressed qualitatively as most calculations rely upon numerical results with larger values of the source-sink separation time, where the size of the excited-state contributions are relatively smaller and the stochastic noise is larger, thus limiting the ability to obtain a controlled, quantitative understanding of the excited state contributions to the correlation function.

In order to address these issues, we have generated results with a large number of short to intermediate values of the source-sink separation time (13 values for $\tsep\approx0.18 - 1.22$~fm) on an a09m310 ($a\approx0.09$~fm, $m_\pi\approx310$~MeV) ensemble.
The large number of $\tsep$ values with precise numerical results allows us to include up to five states in the correlation function analysis while performing a variety of data cuts, as discussed in detail in~\secref{sec:lattice_calc}.

We focus our analysis and discussion on $g_A$, the nucleon matrix element of the axial current in the forward limit, as this matrix element has proved to be one of the most challenging regarding control of excited state contamination.
The first LQCD calculation of $g_A$ with relatively light dynamical quarks ($m_\pi \gtrsim 350$~MeV) appeared in 2005~\cite{Edwards:2005ym}, resulting in a value that had a 7\% statistical uncertainty and agreed with the experimental value after extrapolation to the physical pion mass.
This led the community to anticipate $g_A$ would soon become a precision benchmark quantity for LQCD.

However, subsequent calculations confounded these expectations with the results remaining roughly independent of the pion mass (and below the physical value) or, worse, trending away from the physical value as the pion mass was reduced~\cite{Yamazaki:2009zq,Bratt:2010jn,Alexandrou:2010hf}.
It was speculated that the issue might be due to finite volume corrections which were much larger~\cite{Jaffe:2001eb,Cohen:2001bg,Yamazaki:2008py} than predicted by chiral perturbation theory ($\chi$PT)~\cite{Beane:2004rf}.
As groups investigated the sensitivity of the extracted matrix elements as a function of $\tsep$, and added an excited state in the fit model for the correlation function, it became clear that the dominant unresolved issue was contamination from excited states~\cite{Green:2011fg,Dinter:2011sg,Capitani:2012gj,Green:2012ud,Bhattacharya:2013ehc,Bali:2013nla}.%
%-------------------------------------------------------------------------------
% FOOTNOTE
\footnote{It should be noted that there are still relatively large and important finite volume corrections that must be accounted for to achieve percent-level control of the nucleon matrix elements~\cite{Harris:2019bih,Lutz:2020dfi}.}
%-------------------------------------------------------------------------------
After this, a number of calculations were performed~\cite{Capitani:2012gj,Horsley:2013ayv,Bali:2014nma,Abdel-Rehim:2015owa,Bhattacharya:2016zcn,Alexandrou:2017hac,Capitani:2017qpc,Berkowitz:2017gql,Chang:2018uxx,Berkowitz:2018gqe,Walker-Loud:2019cif,Gupta:2018qil,Hasan:2019noy,Alexandrou:2019brg,Park:2021ypf} that were in agreement with the physical value of $g_A$~\cite{Zyla:2020zbs,PhysRevLett.56.919,YEROZOLIMSKY1997240,LIAUD199753,Mostovoi:2001ye,PhysRevLett.100.151801,Mund:2012fq,Mendenhall:2012tz,Markisch:2018ndu,Beck:2019xye}.
Following the computation with the Feynman-Hellmann method described in Ref.~\cite{Bouchard:2016heu}, other groups have also utilized a larger number of $\tsep$ values at the physical pion mass~\cite{Hasan:2019noy,Alexandrou:2019brg} and found an improved understanding of excited states.\footnote{To the best of our knowledge, the use a large number of $\tsep$ values was first advocated for in Refs.~\cite{Detmold:2011bp,Detmold:2012ge}, which demonstrated the benefit with LQCD calculations of heavy-hadron axial matrix elements using five $\tsep$ values.}

Despite this progress, there remains some tension in the literature, in particular between our results~\cite{Berkowitz:2017gql,Chang:2018uxx,Berkowitz:2018gqe,Walker-Loud:2019cif} and those from the PNDME Collaboration~\cite{Bhattacharya:2016zcn,Gupta:2018qil}, both of which are the only results to utilize three (CalLat) or four (PNDME) lattice spacings and physical pion masses.
Both sets of results were generated with mixed actions that use the $N_F=2+1+1$ Highly Improved Staggered Quark (HISQ) action~\cite{Follana:2006rc} in the sea-quark sector generated by the MILC collaboration~\cite{Bazavov:2012xda} and, in the former case, also by the CalLat Collaboration~\cite{Miller:2020xhy,Miller:2020evg}.
The CalLat results are generated with a M\"obius~\cite{Brower:2012vk} Domain-Wall~\cite{Kaplan:1992bt,Shamir:1993zy,Furman:1994ky} Fermion (MDWF) valence action~\cite{Berkowitz:2017opd} and are computed with $a\approx\{0.09, 0.12, 0.15\}$~fm lattice spacings while the PNDME results are generated with a tadpole-improved~\cite{Lepage:1992xa} clover-Wilson valence action with $a\approx\{0.06, 0.09, 0.12, 0.15\}$~fm.

In Ref.~\cite{Gupta:2018qil}, it was shown that the tension between the CalLat and PNDME results is driven by the PNDME results on the $a\approx0.06$~fm ensembles, which tend to pull the final result to a smaller value, suggestive that the discrepancy may be a discretization effect.
However, it was pointed out that the lever-arm in values of $\tsep$ between the smallest and largest source-sink separation times was the smallest on the $a\approx0.06$~fm ensembles, and that the high-correlation between neighboring time-slices on these fine ensembles makes them more susceptible to correlated fluctuations~\cite{andre:latt2019}.
Further, subsequent analysis by PNDME demonstrated an under-reported excited-state fitting systematic in their results which seems to alleviate the tension~\cite{Jang:2019vkm}.

Another difference between the CalLat and PNDME results is that the PNDME results rely on the more common fixed source-sink separation method while the CalLat results utilize a variant of the summation method~\cite{Maiani:1987by} which can be derived with the Feynman-Hellmann Theorem~\cite{Bouchard:2016heu}.
As was shown in Ref.~\cite{Capitani:2012gj}, the summation method suppresses excited states more than the standard fixed source-sink separation method, and as we will discuss in some detail in \secref{sec:spectral_decomp}, the Feynman-Hellmann derived correlator suppresses excited states even more than the summation method.
The excited state systematic uncertainty, therefore, deserves more scrutiny.  This has been recognized by the community, illustrated by the focus on excited state contamination in the most recent review on nucleon structure from the annual lattice field theory symposium~\cite{Ottnad:2020qbw}.

The challenge of controlling calculations of $g_A$ and other matrix elements has inspired a series of papers aimed at understanding the excited state contamination by utilizing chiral perturbation theory~\cite{Tiburzi:2009zp,Bar:2012ce,Bar:2015zwa,Tiburzi:2015tta,Bar:2016uoj,Bar:2017kxh}.
This work has led to further ideas to try to improve the calculation of the nucleon axial form-factor~\cite{Bali:2019yiy,Bar:2019igf,Jang:2019vkm} as well as an earlier study showing the importance of excited states and consistency of the partially conserved axial current (PCAC) relation from LQCD~\cite{Liang:2018pis}.
It is worth noting, however, that $\chi$PT predicts that the excited state contributions should shift the correlation function above its asymptotic value, while numerical results from all calculations show that the ground state limit is approached from below, and thus, there is a significant discrepancy between this theoretical prediction and the numerical data.
While there are some significant indications that $SU(2)$ baryon chiral perturbation theory, without explicit delta degrees of freedom, is not a converging expansion even at the physical pion mass~\cite{WalkerLoud:2008bp,WalkerLoud:2008pj,Walker-Loud:2013yua,Walker-Loud:2014iea,Chang:2018uxx,Walker-Loud:2019cif,Drischler:2019xuo}, one might anticipate that the predictions from chiral perturbation theory should be at least qualitatively correct.
For $g_A$, it is known that the inclusion of explicit delta degrees of freedom leads to a partial cancellation of the NLO corrections~\cite{Hemmert:2003cb,Procura:2006gq}, which can be understood from large $N_c$ arguments~\cite{Flores-Mendieta:2006ojy,CalleCordon:2012xz}.  The delta degrees of freedom also lead to competing finite volume corrections as compared to the virtual nucleon-pion loops~\cite{Beane:2004rf}.
A more careful investigation of such effects is warranted.

In this work, we take a data-driven approach and ask, given a large dataset, what can we learn about the excited state contamination of the nucleon axial-vector three-point function?
We begin with a summary of the spectral representation of the three point functions in~\secref{sec:spectral_decomp}, then we turn to our numerical results and analysis in~\secref{sec:lattice_calc}.
We offer some observations and conclusions in~\secref{sec:conclusions}, and we present extensive details of our results and analysis in the appendices.

\section{\label{sec:spectral_decomp} Spectral Decomposition}
Lattice QCD calculations are performed in Euclidean space in a mixed time-momentum basis.
In this paper, we focus on the forward matrix element at zero momentum.
Most LQCD calculations are performed with a local creation operator and a momentum-space annihilation operator.
With such a setup, the relevant two-point correlation function at zero momentum and time separation $\tsep$ is given by
\begin{align}\label{eq:2pt}
C_2(\tsep) &= \sum_\mathbf{x} \langle \Omega | N(\tsep,\mathbf{x}) N^\dagger(0,\mathbf{0}) | \Omega \rangle
\nonumber\\&=
    \sum_{n=0}^{\infty} |z_n|^2 e^{-E_n \tsep}
\nonumber\\&=
    |z_0|^2 e^{-E_0\tsep}\left[ 1
        +\sum_{n=1}^\infty |r_n|^2 e^{-\D_{n0}\tsep}
    \right]\, .
\end{align}
In this expression, we assume that the overlap factors, $z_n=\langle\O|N|n\rangle$, used to create ($N^\dagger$) and annihilate ($N$) the states with quantum numbers of the nucleon from the vacuum ($|\O\rangle$) are conjugate to each other.\footnote{We are using the non-relativistic normalization $\langle n | m\rangle = \d_{nm}$ and $\mathds{1} \equiv \ket{\O}\bra{\O} +\sum_{n=0}^\infty \ket{n}\bra{n}$.
We assume that the contributions from the thermal terms that are exponentially suppressed from the full temporal extent are sufficiently suppressed that they can be ignored.}
In the last line, we have defined the energy splitting and ratio of overlap factors:
\begin{align}
&\D_{mn} = E_m - E_n\, ,&
&r_n = \frac{z_n}{z_0}\, .&
\end{align}
The parameterization of \eqnref{eq:2pt} recasts all excited-state parameters with respect to the ground-state, and yields a more universal set of excited-state distributions, simplifying the estimation of their starting values in a frequentist minimization or prior distribution in a Bayesian minimization~\cite{Horz:2020zvv}.

\subsection{Three-point correlators\label{sec:three-point}}

The matrix elements of interest (with Dirac and isospin structure $\G$) are determined through an analysis of three-point correlation functions which are also computed in a mixed time-momentum basis.
The most common strategy is to use a sink with fixed definite spatial momentum at $t=\tsep$ with a current insertion ($j_\G$) at $\t$, between the source ($N^\dagger$) at $t=0$ and the sink ($N$).
In the limit of zero momentum and zero-momentum transfer, the three-point function is given by
\begin{align}\label{eq:3pt_spec}
C_\G(\tsep,\t)
    &= \sum_{\mathbf{y,x}}
	\langle \Omega| N(\tsep,\mathbf{y})\, j_\G(\t,\mathbf{x})\, N^\dagger(0,\mathbf{0}) | \Omega\rangle
\nonumber\\&
	= \sum_n |z_n|^2\ g^\G_{nn}\ e^{-E_n \tsep}
\nonumber\\&\phantom{=}
    +2\sum_{n<m} z_n z^\dagger_m \ g^\G_{nm}\ e^{-(E_n +\frac{\D_{mn}}{2}) \tsep}
\nonumber\\&\phantom{=}\qquad\times
        \cosh \left[ \D_{mn}\left(\t -\frac{\tsep}{2}\right)\right]\, ,
\end{align}
where the matrix elements of interest are given by
\begin{align}
g^\G_{mn} = \langle m| j_\G |n\rangle \, .
\end{align}
In this limit, the correlation functions are real, so $z^\dagger_n = z_n$ and $g^\G_{nm} = g^\G_{mn}$, which are used to simplify \eqnref{eq:3pt_spec}.

Typically one determines the ground state matrix element, $g^\G_{00}$, by performing the computation for 1, 2, 3 or sometimes 4 values of $\tsep$ in the range $\tsep\approx1-1.5$~fm while utilizing local three-quark interpolating fields for the creation and annihilation operators.
The correlation functions are then analyzed using models with zero, one, or more excited states~\cite{Capitani:2012gj,Horsley:2013ayv,Bali:2014nma,Abdel-Rehim:2015owa,Bhattacharya:2016zcn,Alexandrou:2017hac,Capitani:2017qpc,Berkowitz:2017gql,Chang:2018uxx,Berkowitz:2018gqe,Walker-Loud:2019cif,Gupta:2018qil,Hasan:2019noy,Alexandrou:2019brg,Park:2021ypf}, with some analysis using up to four states~\cite{Yoon:2016jzj,Rajan:2017lxk,Jang:2019vkm}.
There are several challenges and shortcomings with this strategy, which we will summarize.
We note the LHP and ETM Collaborations have performed calculations with significantly more values of $\tsep$ than is common~\cite{Hasan:2019noy,Alexandrou:2019brg}.

The ground state (gs) matrix element $g^\G_{00}$ is typically determined by constructing the ratio correlation function
\begin{equation}\label{eq:ratio}
R_\G(\tsep,\t) = \frac{C_\G(\tsep,\t)}{C_2(\tsep)}\, ,
\end{equation}
which gives the ground state matrix element in the limit
\begin{equation}
\lim_{\tsep\rightarrow\infty} R_\G(\tsep,\t\approx\tsep/2) = g^\G_{00}\, .
\end{equation}

The energies and overlap factors can be constrained from the two-point function, leaving the three-point function to constrain the matrix elements $g^\G_{nn}$ and $g^\G_{nm}$.\footnote{Some groups fix the overlap factors and energies from the two-point correlation function and then pass the central or correlated values into a three-point function analysis.
This strategy can be particularly problematic, as emphasized in Ref.~\cite{Jang:2019vkm}, because the two-point functions generated from purely local three-quark interpolating operators are not sufficiently constraining of the excited state spectrum, which can lead to significant differences in the extracted ground state matrix elements.
Instead, we perform a simultaneous fit to the two- and three-point functions.
}
From \eqnref{eq:3pt_spec}, one observes that only the transition terms ($n\neq m$) are sensitive to the current insertion time $\t$.
To isolate the ground state matrix element, the minimum number of values of $\tsep$ required is at least one greater than the number of states used in the analysis in order to have a 1-degree-of-freedom fit.
For example, a two-state fit requires the determination of $g^\G_{00}$, $g^\G_{11}$ and $g^\G_{01}$: the latter can be constrained from the $\t$ dependence leaving the remaining $\tsep$ dependence to constrain the two former matrix elements.
In this minimal scenario, computations which utilize three values of $\tsep$ (or less) are not able to perform a systematic study on omitting values of $\tsep$, or changing the number of states used in the analysis.
This prohibits a verification that the full uncertainty on $g^\G_{00}$ associated with excited state contamination has been correctly captured.
In other words, we would have to assign an unquantified systematic uncertainty to the ground state matrix element.

Another significant challenge is that lattice computations of three-point functions are typically performed for $\tsep\gtrsim1$~fm, which is roughly the time separation when the stochastic noise becomes significant.
Again, because of the limited values of $\tsep$ typically used, one cannot determine if the data at this time is susceptible to a correlated fluctuation or not, which, if present, would cause a bias in the results.
We will return to this point in \secref{sec:lattice_calc}.

To understand the various sources of excited state (es) contamination, we reorder \eqnref{eq:ratio} in a way which deliberately disentangles the different types of excited state contributions.
For the two-point correlation function, it is straightforward to separate the ground state from the excited states
\begin{subequations}
\begin{align}
\label{eq:2pt_gs_es}
C_2(\tsep) &= C_2^{\rm gs}(\tsep) + C_2^{\rm es}(\tsep)\, ,
\\
\label{eq:2pt_gs}
C_2^{\rm gs}(\tsep) &= |z_0|^2 e^{-E_0 \tsep}\, ,
\\
\label{eq:2pt_es}
C_2^{\rm es}(\tsep) &= \sum_{n\geq1} |z_n|^2 e^{-E_n \tsep}\,.
\end{align}
\end{subequations}
For the three-point functions, we denote the $n$-to-$n$ as scattering (sc) states and the $n$-to-$m$ as transition (tr) states such that
\begin{subequations}
\begin{align}
\label{eq:3pt}
C_\G(\tsep,\t) &= C_\G^{\rm gs}(\tsep) + C_\G^{\rm sc}(\tsep) + C_\G^{\rm tr}(\tsep,\t)\, ,
\\
\label{eq:3pt_gs}
	C_\G^{\rm gs}(\tsep) &= |z_0|^2 g^\G_{00} e^{-E_0 \tsep} = g^\G_{00} C_2^{\rm gs}(\tsep)\, ,
\\
\label{eq:3pt_sc}
C_\G^{\rm sc}(\tsep) &= \sum_{n\geq1} |z_n|^2 g^\G_{nn} e^{-E_n \tsep}\, ,
\\
\label{eq:3pt_tr}
C_\G^{\rm tr}(\tsep,\t) &= \sum_{n<m} z_n z^\dagger_m \ 2 g^\G_{nm}\ e^{-(E_n +\frac{\D_{mn}}{2}) \tsep}
\nonumber\\&\qquad\qquad\times
    \cosh \left[ \D_{mn}\left(\t -\frac{\tsep}{2}\right)\right]\, .
\end{align}
\end{subequations}
The ratio correlation function can then be expressed as
\begin{align}\label{eq:ratio_es_separated}
R_\G(\tsep,\t) &=
    \frac{C_\G^{gs}(\tsep) + C_\G^{\rm sc}(\tsep) + C_\G^{\rm tr}(\tsep,\t)}{C_2(\tsep)}
\nonumber\\&=
    g^\G_{00}
    + \frac{C_\G^{\rm sc}(\tsep) - g^\G_{00} C_2^{\rm es}(\tsep)}{C_2(\tsep)}
\nonumber\\&\phantom{=}
    +\frac{C_\G^{\rm tr}(\tsep,\t)}{C_2(\tsep)}
\nonumber\\&=
    g^\G_{00}
    + \frac{\sum_{n\geq1} (g^\G_{nn} - g^\G_{00}) |z_n|^2 e^{-E_n \tsep}}{C_2(\tsep)}
\nonumber\\&\phantom{=}
        +\frac{C_\G^{\rm tr}(\tsep,\t)}{C_2(\tsep)}\, .
\end{align}
Consider the leading excited state contamination to $g^\G_{00}$ arising from $g^\G_{11}$, and $g^\G_{01}$
\begin{align}\label{eq:ratio_limit}
R_\G(\tsep,\t) &\approx
    g^\G_{00} + |r_1|^2(g^\G_{11}-g^\G_{00} ) e^{-\D_{10}\tsep}
\nonumber\\&\phantom{=}
    +2 r_1^\dagger g^\G_{01} e^{-\D_{10}\frac{\tsep}{2}}
    \cosh\left[\D_{10}\left(\t-\frac{\tsep}{2}\right)\right]
\nonumber\\&\phantom{=}
    +\cdots\, ,
\end{align}
where the $\cdots$ includes terms from higher excited states as well as from the first excited state, but further suppressed by extra powers of $\exp(-\D_{10}\tsep)$.
The scattering and two-point excited state contributions, in addition to cancelling against each other for same-sign values of $g^\G_{11}$ and $g^\G_{00}$, are suppressed by the full $\tsep$, $\exp(-\D_{10}\tsep)$.
In contrast, the transition excited states are only suppressed by half the time separation, $\exp(-\D_{10} \frac{\tsep}{2})$, and so they are expected to be the dominant source of excited state contamination.

\subsection{Feynman-Hellmann and Summed Correlators\label{sec:fh_sum}}
Rather than analyzing the three-point functions, one can construct a correlation function in which the current insertion time is summed over, cutting $\tau_c$ timeslices from either end of the correlation function
\begin{align}\label{eq:C3_sum}
\S_\G(\tsep,\t_c) &= \sum_{\t=\t_c}^{\tsep-\t_c} C_\G(\tsep,\t)
\nonumber\\&
    = \sum_n |z_n|^2 (\tsep+1-2\t_c) g^\G_{nn} e^{-E_n \tsep}
\nonumber\\&\phantom{=}
    + 2\sum_{n<m} z_n z^\dagger_m  g^\G_{nm}
        e^{-(E_n +\frac{\D_{mn}}{2}) \tsep}
\nonumber\\&\phantom{=}\quad
        \times
        \frac{\sinh \left[ \frac{\D_{mn}}{2}(\tsep+1-2\t_c)\right]}
            {\sinh \left( \frac{\D_{mn}}{2} \right)}\, .
\end{align}
The original implementation of this idea included a sum over all timeslices, including the contact terms ($\t=0$ and $\t=\tsep$) as well as ``out-of-time-order'' regions ($\t<0$ and $\t>\tsep$)~\cite{Maiani:1987by}.
Given a set of fixed source-sink separation datasets, one is of course free to perform the sum in a variety of ways, e.g. excluding data from the out-of-time-order regions, excluding the source and sink time (setting $\t_c=1$ in \eqnref{eq:C3_sum}) or also cutting time near the source and sink ($\t_c >1$).

The summed correlation function at large $\tsep$
\begin{multline}
\S_\G(\tsep,\t_c) =
|z_0|^2 e^{-E_0\tsep} \bigg\{
\\
    (\tsep+1-2\t_c)\Big[g^\G_{00} + |r_1|^2 g^\G_{11}e^{-\D_{10}\tsep} \Big]
\\
    +r_1^\dagger  g^\G_{01}\frac{
        e^{\frac{\D_{10}}{2}(1-2\t_c)}
        -e^{-\D_{10}(\tsep +\frac{1}{2}-\t_c)}
        }{\sinh(\frac{\D_{10}}{2})}
    \bigg\}
+\cdots\, ,
\end{multline}
can be used to determine the leading excited state contamination.
Note that the ground state and the scattering ($n$-to-$n$) states are relatively enhanced by $\tsep+1-2\t_c$.
The transition matrix elements ($m$-to-$n$) lead to a $\tsep$-independent term and those that depend upon $\tsep$ become exponentially suppressed by the full excited state gap ($e^{-\D_{10}\tsep}$) rather than half the gap, as with the three-point function, \eqnref{eq:ratio_limit}, as noted in Ref.~\cite{Capitani:2012gj}.
Thus, one expects that the excited state contamination of the summed correlation function is smaller than for the standard fixed source-sink separation time three-point correlation function, up to this $\tsep$-independent term.

This summed correlation function has received some attention in the literature~\cite{Gusken:1989ad,Bulava:2011yz,Capitani:2012gj,deDivitiis:2012vs}.
Recently, there have been calculations which utilize a Feynman-Hellmann approach, by performing a computation in the presence of background fields and extracting the matrix elements through the linear response of the spectrum to the background field~\cite{Chambers:2014qaa,Chambers:2015bka,Savage:2016kon}.
In Ref.~\cite{Bouchard:2016heu}, it was shown that the application of the Feynman-Hellmann theorem to the effective mass directly leads to a derivative of the summed correlation function~\cite{Maiani:1987by}, relating the matrix element to the spectrum without the need for an explicit background field.
We call this the Feynman-Hellmann (FH) correlation function\footnote{In the original implementation~\cite{Bouchard:2016heu}, the sum over the current time is over all time-slices, which we denote $\t_c={\rm none}$, as in the original summation method~\cite{Maiani:1987by}.}
\begin{align}\label{eq:fh_corr}
{\rm FH}_\G(\tsep,\t_c,dt) &= \sum_{\t=\t_c}^{\tsep-\t_c} \hspace{-4pt}\frac{
    R_\G(\tsep+dt,\t) - R_\G(\tsep,\t)}{dt}
\nonumber\\&=
    \frac{1}{dt}\left[
        \frac{\S_\G(\tsep+dt,\t_c)}{C_2(\tsep+dt)}
        - \frac{\S_\G(\tsep,\t_c)}{C_2(\tsep)}
    \right],
\nonumber\\
{\rm FH}_\G(\tsep,\t_c) &\equiv {\rm FH}_\G(\tsep,\t_c,dt=1)\, .
\end{align}
Since the FH correlation function is constructed from $\S_\G(\tsep,\t_c)$, it enjoys the larger suppression of excited states, with the leading excited state contamination scaling as $\exp(-\D_{10}\tsep)$ rather than with $\D_{10}/2$ (the $\tsep$ independent pieces exactly cancel in the numerical derivative).
Additionally, the numerical derivative serves to both isolate the ground state, whose contribution grows linearly in $\tsep$, as well as to further suppress the scattering ($n$-to-$n$) and transition ($m$-to-$n$) excited states which do not strongly differ from one timeslice to the next.
This stronger suppression of excited states is what allowed us to utilize earlier Euclidean time data~\cite{Berkowitz:2017gql,Chang:2018uxx,Berkowitz:2018gqe} than is common in the three-point correlation function analysis and to enjoy the benefits of the lower stochastic noise.
We will show this in some detail in \secref{sec:full_results}.

\section{\label{sec:lattice_calc} Lattice Calculation}
\begin{figure*}
\begin{tabular}{cc}
\includegraphics[width=\columnwidth]{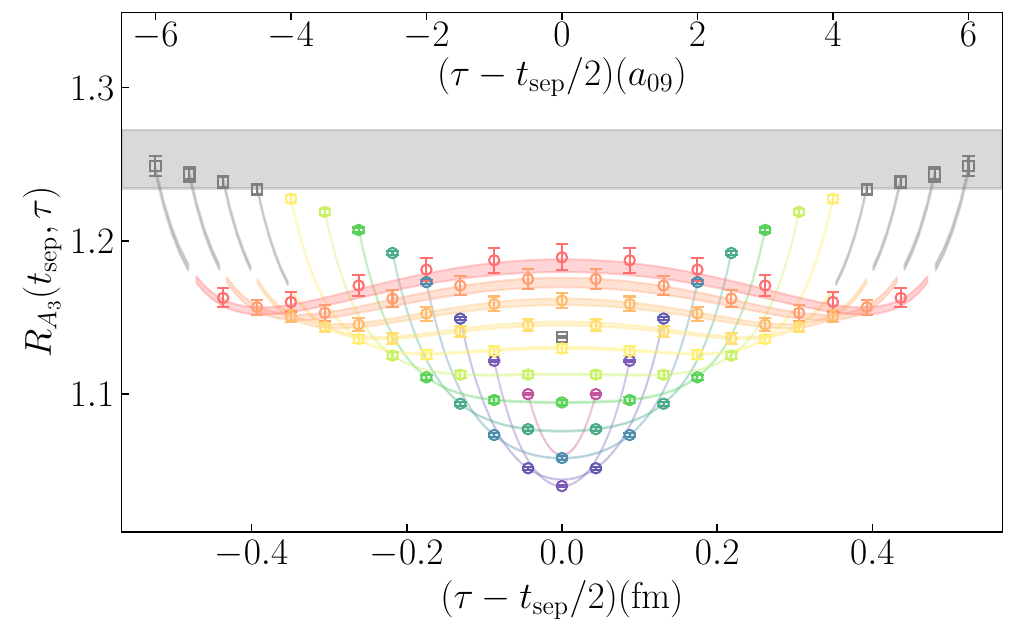}
&\includegraphics[width=\columnwidth]{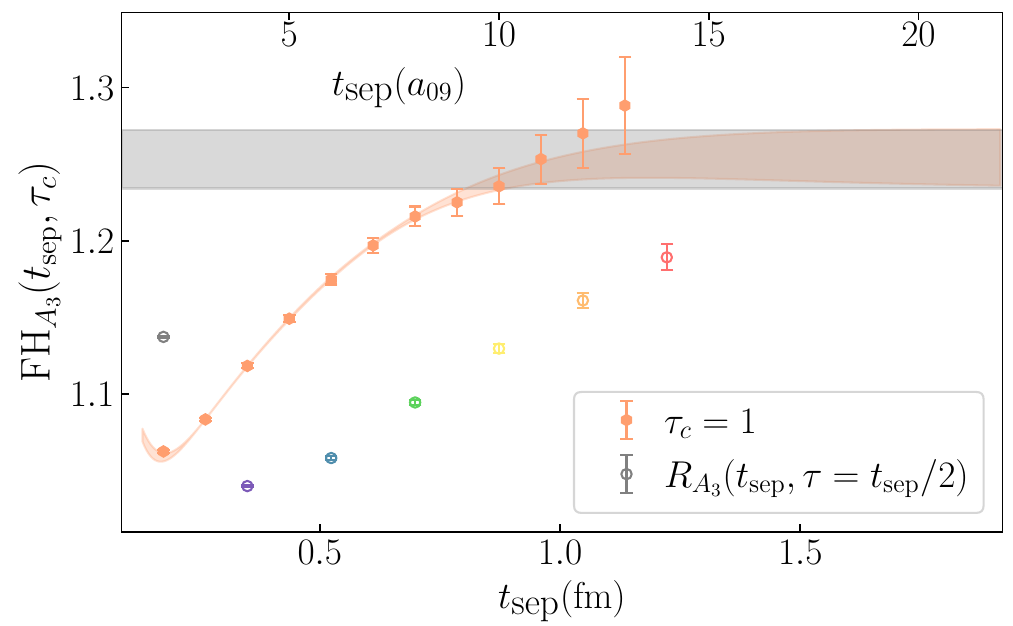}
\\
(a) &(b)
\end{tabular}
\caption{\label{fig:psychedelic_moose}
Left: (Psychedelic Moose/Water Buffalo plot)
We plot the numerical results of $R_{A_3}(\tsep,\t)$ for $\tsep=2$ (the single gray point in the middle) to $\tsep=14$, the top (red) dataset.
In addition, we plot the resulting posterior description of the correlation function from our 5-state fit as the (correspondingly colored) fit bands.
The (gray) squares in the upper left/right of the ``moose antlers'' are not included in the analysis, as indicated by the break in the fit band from the inner region.
The horizontal (gray) band is the ground state matrix element, $\mathring{g}_A$.
Right: The ${\rm FH}_{A_3}(\tsep,\t_c=1)$ numerical data (filled orange symbols) is plotted along with the posterior description of the correlation function from the global analysis.
As a comparison, for even values of $\tsep$, we plot $R_{A_3}(\tsep,\t=\tsep/2)$ with open (colored) symbols (the numcerical data in the middle of the ``moose'').
The time-axis is converted from lattice units (top) to fm (bottom) using our scale-setting~\cite{Miller:2020evg}.
}
\end{figure*}

For the present study, we use results from our MDWF on gradient-flowed HISQ action~\cite{Berkowitz:2017opd} on the a09m310 ensemble, which has a lattice spacing of $a\approx0.09$~fm and a pion mass of $m_\pi\approx310$~MeV.
We use the same parameters as in Refs.~\cite{Berkowitz:2017opd,Berkowitz:2017gql,Chang:2018uxx} except we change the parameters that go into the \texttt{GAUGE\_COVARIANT\_GAUSSIAN} smearing routine in \chroma~\cite{Edwards:2004sx} to $\sigma_{\rm smr} = 3.5$ and $N_{\rm smr} = 45$.

We generate 16 sources per gauge configuration on 784 configurations.
We generate the three-point functions using a sequential-propagator through the sink at 13 values of $\tsep$,
\begin{equation}
\tsep/a_{09} \in \{2,3,4,5,6,7,8,9,10,11,12,13,14\}\, .
\end{equation}
Our sources and sinks are generated with a local three-quark interpolating field using only the upper-spin components of the quark field in the Dirac-Pauli basis (lower components for the negative parity states), which gives the largest overlap onto the ground state of the nucleon at rest~\cite{Basak:2005aq,Basak:2005ir}.
In \appref{app:discrete_symmetry}, we present further details of our computation, including the cost-benefit analysis of improving the stochastic sampling by combining 8 coherent sinks~\cite{Bratt:2010jn} for each sequential propagator, the use of spin up and spin down sources and sinks (versus utilizing a spin-projector that isolates one of the spin states) and the use of time-reversed negative parity correlators.

\subsection{\label{sec:full_results} Full results}

We begin with a presentation of our final results which come from a fully-correlated Bayesian constrained curve-fit~\cite{Lepage:2001ym} with a five-state model to describe
\begin{align*}
&C_2(\tsep)\, ,& &R_{A_3}(\tsep,\t)\, ,& &R_{V_4}(\tsep,\t)\, ,&
\\
&& &{\rm FH}_{A_3}(\tsep,\t_c=1)\, ,& &{\rm FH}_{V_4}(\tsep,\t_c=1)\, .&
\end{align*}
The final result is obtained with all values of $\t$ between the source and sink time, $\t=[1,\tsep-1]$.
For $R_\G(\tsep,\t)$, the results are symmetrized about $\t=\tsep/2$ and half the data (plus $\t=\tsep/2$ point for even values of $\tsep$) is used in the analysis.

In the left panel of \figref{fig:psychedelic_moose} (the ``Psychedelic Moose''), we plot the numerical results for the ratio of the three-point function generated with the $A_3=\bar{q}\g_3\g_5 \t_3 q$ current, divided by the two-point function at the given value of $\tsep$ (cfr. \eqnref{eq:ratio}).
We also plot the resulting posterior description with our 5-state model.
The fit quality is good and visually one can see that the model accurately describes the numerical results over the full range of $\tsep$ and $\t$ used in the analysis.
The horizontal (gray) band is the value of the ground state matrix element $\mathring{g}_A$.
In the right panel we plot the ${\rm FH}_{A_3}(\tsep,\t_c=1)$ data that are used in the global fit as well as the resulting posterior distribution of this correlation function.

\begin{figure}
\includegraphics[width=\columnwidth]{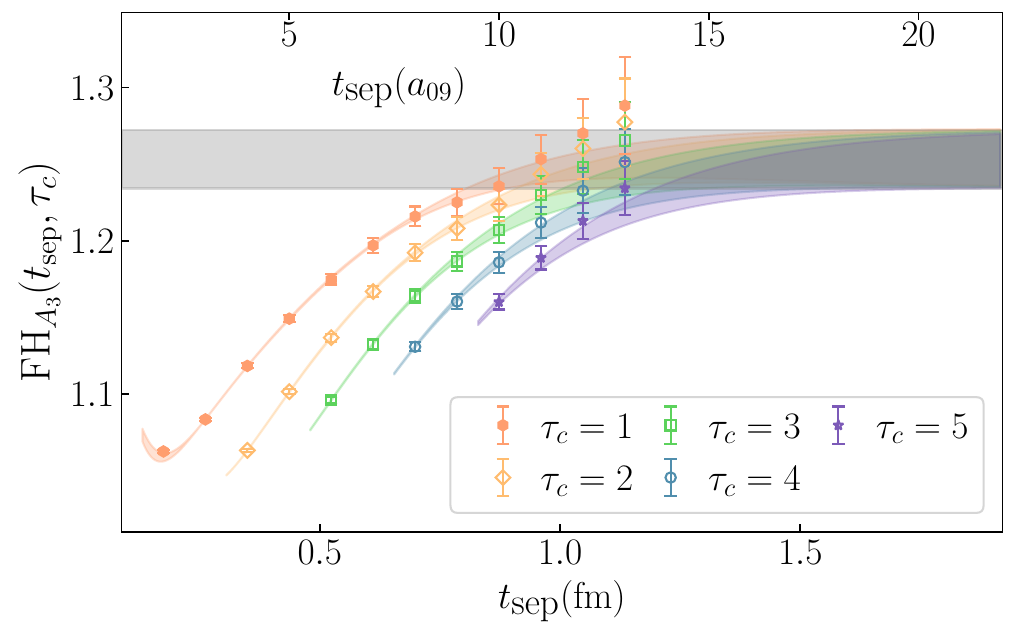}
\caption{\label{fig:tau_cut}
We plot the numerical ${\rm FH}_{A_3}(\tsep,\t_c)$ data for various values of $\t_c$ as well as the posterior reconstruction of these correlation functions from the global analysis that uses $\t_c=1$.
}
\end{figure}

In \figref{fig:tau_cut}, we explore the ${\rm FH}_{A_3}(\tsep,\t_c)$ data as the number of data near the source and sink time are cut from the sum over current insertion time denoted by increasing $\t_c$.
The posterior fit bands are from the global analysis that uses $\t_c=1$.
As $\t_c$ is increased, for a fixed $\tsep$, one observes that the excited state contribution becomes larger.
This can be understood by looking at the leading excited state contribution to ${\rm FH}_\G(\tsep,\t_c)$ that depends upon $\t_c$
\begin{equation*}
{\rm FH}_\G(\tsep,\t_c) \ni
    e^{\D_{10}(\t_c+\frac{1}{2})}
    r_1^\dagger g_{01}^\G
    \frac{e^{-\D_{10}\tsep} - e^{-\D_{10}(\tsep+1)}}{\sinh \frac{\D_{10}}{2}}.
\end{equation*}
As is evident from \figref{fig:psychedelic_moose}, the transition matrix element $g_{01}^{A_3} < 0$; an observation common to all LQCD calculations of $g_A$.
Therefore, the leading excited state contamination that depends upon $\t_c$ is negative and grows exponentially with increasing $\t_c$, consistent with the results.

In order to have confidence in our final results for $m_N=E_0$, $\mathring{g}_A$ and $\mathring{g}_V$,\footnote{With our action, the axial-vector and vector renormalization factors are equal to $10^{-4}$ so the renormalized axial charge is given by the ratio of the bare matrix elements, $g_A = \mathring{g}_A / \mathring{g}_V$~\cite{Chang:2018uxx}.}
as well as the ability to quantify and control the excited state contamination, we discuss our analysis strategy and the stability of our results under model variation and data truncation.

\subsection{\label{sec:analysis_strategy} Analysis strategy}
\begin{figure*}
\includegraphics[width=\textwidth]{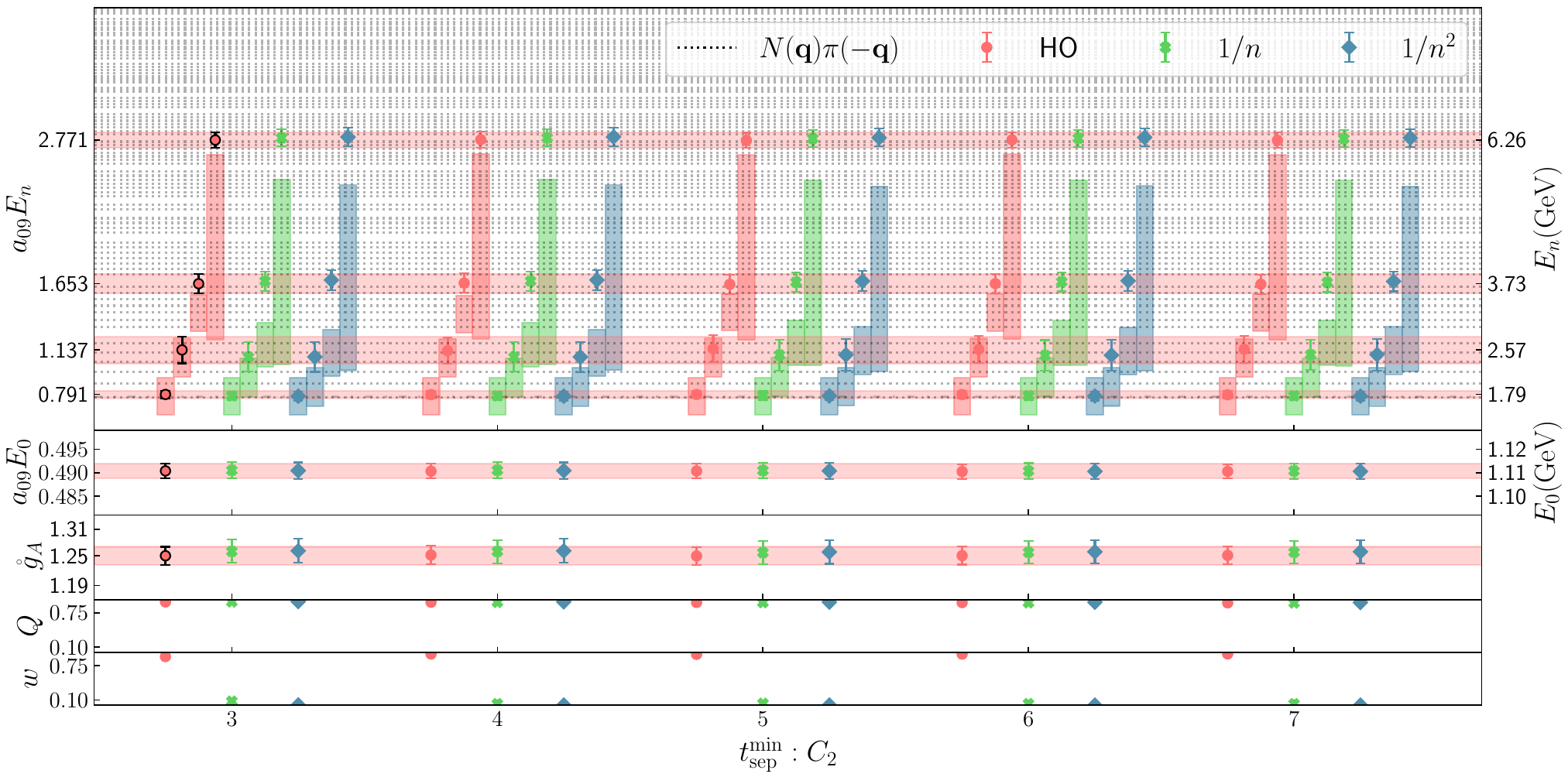}
\caption{\label{fig:excited_state_model}
Sensitivity of the extracted spectrum and $\mathring{g}_A$ on the model of excited states used as a function of $\tsep^{\rm min}$ in the $C_2(\tsep)$ correlation function.
For a given $\tsep^{\rm min}$, from left to right, offset for visual clarity, we plot the prior (vertical (colored) box) and posterior values (filled markers) of $\mathring{g}_A$, $E_0$ and $E_n$ for the harmonic-oscillator (HO), inverse $n$ ($1/n$) and inverse $n^2$ ($1/n^2$) excited state models, \eqnref{eq:spectrum_parameterization}.
The prior width on $\mathring{g}_A$ and $E_0$ are larger than the displayed $y$-limits, and thus not shown.
The horizontal light (red) bands (to guide the eye) are from the chosen fit denoted with the black border on the (colored) markers.  The horizontal dashed lines denote the non-interacting energy levels of the $P$-wave $N\pi$ states.
See the text for more details.
}
\end{figure*}

We have two goals with this work:
\begin{enumerate}[leftmargin=*]
  \item Identify the ground state mass and matrix elements, $\mathring{g}_A$ (and $\mathring{g}_V$), with a complete systematic uncertainty;
  \item Obtain a quantitative understanding of the excited state contamination of the correlation functions used in the analysis.
\end{enumerate}
The former is possible without the latter through a demonstration that the extracted ground state mass and matrix elements are invariant under modifications of the fit model (i.e. the number of excited states and the model of their mass-gap) and truncations in the time range used in the analysis.

A robust identification of the excited state spectrum and matrix elements would require the use of a variational basis that includes multi-hadron operators~\cite{CP-PACS:2007wro,Gockeler:2008kc,Budapest-Marseille-Wuppertal:2010gis,Feng:2010es,Lang:2011mn,Dudek:2012xn,Lang:2012db,Wilson:2015dqa}.
For this reason, our second goal is to obtain a quantitative understanding of the sum of all excited state contributions, and in particular, to quantify how they contaminate the ground state values.
To achieve this goal it is essential that the fit model accurately describes the correlation function over the full range of time separations without over-fitting the data.

An important point to note is that, with only the local three-quark interpolating operators, it is anticipated that the excited states determined in the analysis will be linear combinations of the true eigenstates of the system.
Therefore, the excited state spectrum determined from the two-point function may not be unique, and may be different than that determined from a simultaneous fit with the three-point functions.
Significant attention was given to this issue in Ref.~\cite{Jang:2019vkm}: when the excited states used in the analysis of the three-point function were priored to follow the anticipated $P$-wave $N(\mathbf{q})\pi(-\mathbf{q})$ (non-interacting) energy levels, the value of the ground state matrix element was found to shift significantly as compared to when the spectrum was taken from the analysis of the two-point correlation functions.
In both analyses, the authors did not perform a simultaneous analysis of the two-point and three-point functions, but instead used the excited state spectrum determined from either the two-point function of the $A_4$ three-point function to inform priors for the analysis of the three-point function.

We take a different viewpoint.
Because the interpolating operator basis is not sufficient to uniquely identify the spectrum, and the value of the ground state matrix elements can change significantly based upon different models of excited states, which all have acceptable $\chi^2/{\rm dof}$ in the analysis, it is critical to perform a minimization of all correlation functions simultaneously.  Only with this global minimization is it possible to infer how similar or dissimilar the various analysis strategies are.

In \appref{app:analysis_details}, we provide a detailed description of the analysis and the sensitivity of the ground state posteriors under variations in the number of states used, the model of excited states and the fit ranges in $\tsep$ and $\t$.
Here, we summarize the findings of this study.

We begin with a discussion of the model for the excited states.
The lowest-lying excited states consist roughly of either a nucleon-pion in a relative P-wave or a nucleon with a two-pion excitation.
With $m_\pi L\approx4$ the non-interacting energy levels of these two states are nearly identical and so a calculation without multi-hadron operators can not distinguish them.  Therefore, they are treated as a single excited state.
We then use three different models for the spectrum of excitations and parametrize the energy gaps $\D E_n$, where $n$ indicates the $n^{th}$ state, as:
\begin{align}\label{eq:spectrum_parameterization}
\textrm{$n^{th}$ energy level}   &&& E_n = E_0 + \sum_{l=1}^n \D E_l\, , \nonumber\\
\textrm{harmonic oscillator (HO)}&&& \D E_n = 2 m_\pi\, , \nonumber\\
\textrm{inverse $n$ ($1/n$)}     &&& \D E_n = 2m_\pi / n\, , \nonumber\\
\textrm{inverse $n^2$ ($1/n^2$)} &&& \D E_n = 2m_\pi / n^2\, .
\end{align}
For each model, for each level $n<N_{\rm max}$, we set the prior as $\D \tilde{E_n} = (\D E_n, m_\pi)$ where the first entry is the prior mean and the second is the prior width.  For the highest state, which we expect to be a ``garbage can'', we set the prior width to be $5m_\pi$.
We observe two important facts when we vary the excited state model:
\begin{enumerate}[leftmargin=*]
\item The ground state posteriors are insensitive to the model used;
\item The excited state posteriors are mostly insensitive to the model used, even when the posterior is in significant conflict with the prior;
\end{enumerate}
both strong indicators that the extracted energy levels are dictated by the numerical data and not the priors used.
In \figref{fig:excited_state_model}, we show the sensitivity of $\mathring{g}_A$ and the spectrum on the excited state model versus the $\tsep^{\rm min}$ of the two-point correlation function (in all cases for the full five-state, five-correlation function analysis).
The lowest panel is the relative weight determined from the three models at a given $\tsep^{\rm min}$
\begin{equation}\label{eq:w-logGBF}
w_{\rm i} = \frac{ e^{{\rm logGBF}_i}}{\sum_{j\in{\rm models}} e^{{\rm logGBF}_j}}\, ,
\end{equation}
where logGBF is the log of the Gaussian Bayes Factor.
The next panel gives the $Q \in [0,1]$ value which is a measure of the fit-quality.
For the values of $\mathring{g}_A$ and $E_n$, the light (red) horizontal band is the value of these quantities from the chosen fit from the HO model at $\tsep^{\rm min}=3$, to guide the eye, which is also denoted by the markers with a black border.
The vertical bars represent the prior width for the given quantity and are aligned in the same vertical column as their corresponding posterior values (for $\mathring{g}_A$ and $E_0$, the prior widths are larger than the displayed $y$-limits and so they are not shown).\footnote{The energy spectrum is priored with a series of ordered energy splittings.  Therefore, to construct the priors shown in \figref{fig:excited_state_model}, we plot $\tilde{E}_n = \hat{E}_0 + \sum_{l=1}^{n-1} \D \hat{E}_l + \D \tilde{E}_n$ where the priors are denoted with a tilde and the posteriors are denoted with a hat.}
Finally, the horizontal dashed lines give the non-interacting $P$-wave $N(\mathbf{q})\pi(-\mathbf{q})$ energy levels, which are quite dense.  The lowest non-interacting $N(\mathbf{q+p})\pi(\mathbf{-q})\pi(\mathbf{-p})$ level where all hadrons are at rest is nearly degenerate with the lowest $N(\mathbf{q})\pi(-\mathbf{q})$ level.
We further note that:
%In particular,
\begin{enumerate}[leftmargin=*]
\item The extracted posteriors are consistent between models which all have a high fit-quality, even when the posterior value is in tension with the prior.  This is a strong indicator the extracted spectrum is highly constrained by the numerical data and not the priors, even for the high-lying energy levels.

\item The first excited state is consistent with the lowest lying $N\pi$ state.
%(which is nearly degenerate with the $N\pi\pi$ state).
The higher lying states have an uncertainty that spans several anticipated energy levels, indicating that they are likely a linear combination of eigenstates.
\end{enumerate}
In \appref{app:spec_g_consistency}, we show that while the two-point functions are not able to constrain the excited states on their own, when analyzed in combination either the $V_4$ or $A_3$ correlation functions (or both), the extracted spectrum becomes more precise and very stable against variations in the fit model.
This observation supports our conclusion that it is important to perform a global, fully correlated analysis to obtain a reliable determination of the ground state parameters.  A similar observation and conclusion has been made in Ref.~\cite{Harris:2019bih} and reviewed in Ref.~\cite{Ottnad:2020qbw}.

\begin{figure*}
\begin{tabular}{@{\hskip -0.05in}c@{\hskip -0.05in}c@{\hskip -0.05in}c}
\includegraphics[width=0.35\textwidth]{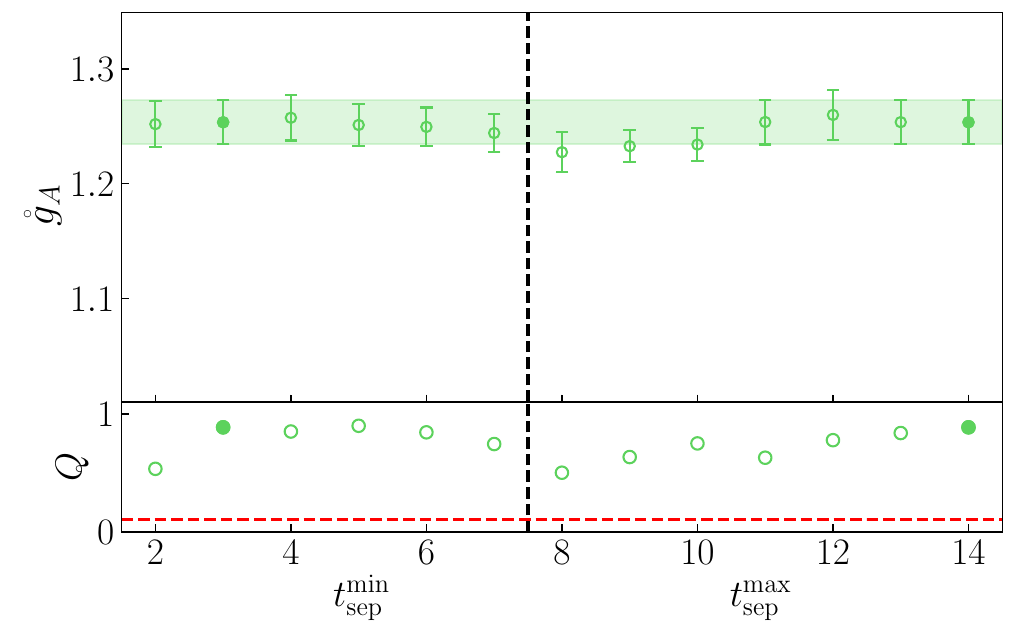}&
\includegraphics[width=0.35\textwidth]{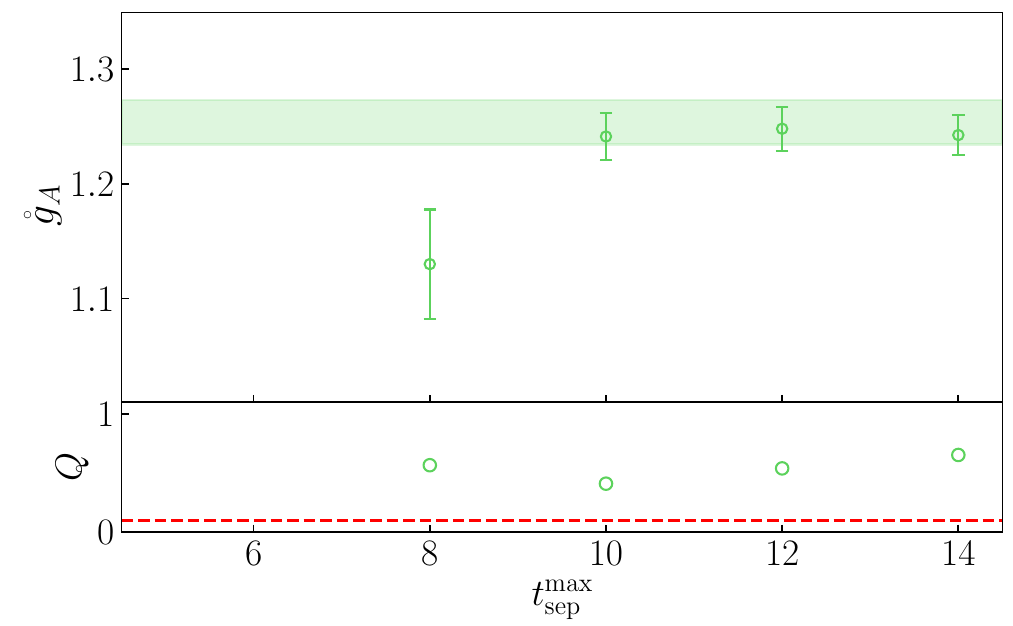}&
\includegraphics[width=0.35\textwidth]{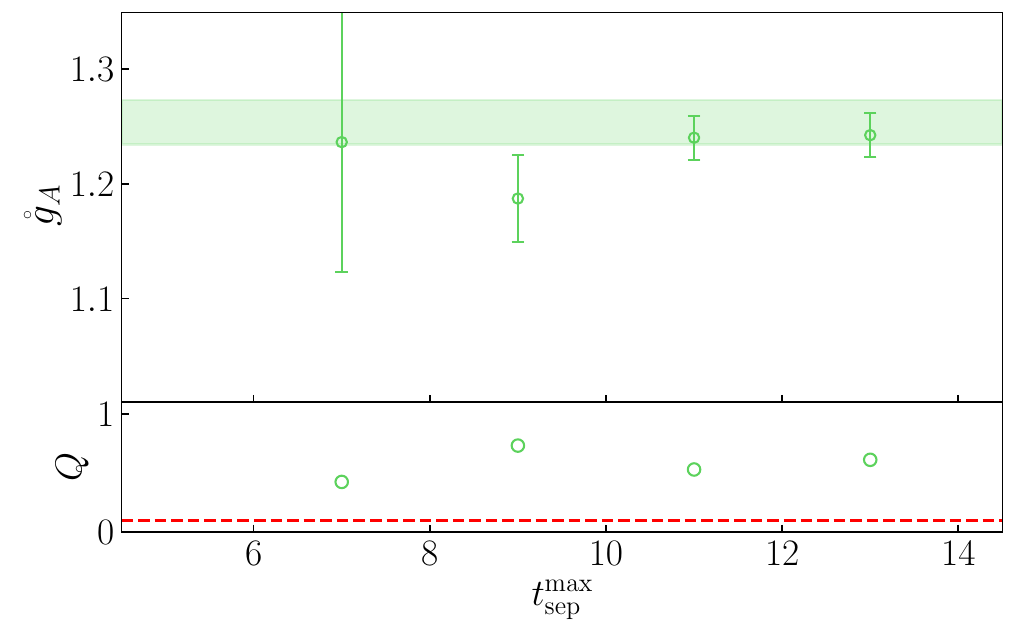}
\\
(a): $\tsep^{\rm min}/\tsep^{\rm max}$ sensitivity&
(b): even $\tsep$ only&
(c): odd $\tsep$ only
\end{tabular}
\caption{\label{fig:stability}
Stability of the determination of $\mathring{g}_A$ under data truncation.
The horizontal band is the value of $\mathring{g}_A$ determined from the full analysis, to guide the eye.
The bottom part of each panel represents the fit quality $Q \in [0,1]$.
The left panel shows the stability as we increase $\tsep^{\rm min}$ while holding $\tsep^{\rm max}=14$ to the left of the vertical dashed line and, similarly, the result as $\tsep^{\rm max}$ is increased with $\tsep^{\rm min}=2$ to the right of the dashed line.
The middle and right panels show the $\tsep^{\rm max}$ stability when only even or odd values of $\tsep$ are used in $R_\G(\tsep,\t)$ and ${\rm FH}_\G(\tsep,\t_c=1,dt=2)$.
}
\end{figure*}

Next, we examine the stability of $\mathring{g}_A$ as we remove data from the fit (all fits still use the set of five correlation functions).
We examine the sensitivity as we increase $\tsep^{\rm min}$ in the analysis, as we reduce $\tsep^{\rm max}$ and also through using only even or odd values of $\tsep$ in the three-point functions.
For each choice of which values of $\tsep$ to use, FH$_\G(\tsep,\t_c=1,dt)$ is constructed from the set of $R_\G(\tsep,\t)$.  When we use consecutive values of $\tsep$, we take $dt=1$ in FH$_\G(\tsep,\t_c=1,dt)$.  When we use only even or odd values of $\tsep$, we take $dt=2$.

The left panel of \figref{fig:stability} shows $\mathring{g}_A$ as a function of the minimum and maximum value of $\tsep$ used in the analysis.
When $\tsep^{\rm min}$ is varied, $\tsep^{\rm max}$ is held fixed at 14.
When $\tsep^{\rm max}$ is varied, $\tsep^{\rm min}$ is fixed at 2.
In the middle panel of \figref{fig:stability}, we show the value of $\mathring{g}_A$ as a function of $\tsep^{\rm max}$ when only even values of $\tsep$ are used in the analysis of $R_\G(\tsep,\t)$ and ${\rm FH}_\G(\tsep)$.
The right panel is the same as the middle panel but we only use the odd, rather than the even, values of $\tsep$ in the analysis.

When the three-point functions are computed with a sufficiently large number of $\tsep$ values, and the two-point and three-point functions are analyzed in a fully correlated global minimization, the ground state parameters are very stable under data truncation and the excited state model.
Of note, it is sufficient to use every other value of $\tsep$ at this $a\approx0.09$~fm lattice spacing.

\subsection{\label{sec:es_breakdown} Excited State Breakdown}
Given a model that is demonstrated to accurately describe the correlation functions over the full range of $\tsep$ and $\t$ used in the analysis, we can separate the various sources of excited state contamination into the ``scattering'' ($n$-to-$n$) and ``transition'' ($m$-to-$n$) sources as well as those arising from the excited states of the two-point function (see~\secref{sec:three-point}).
While we can accurately describe these various sources of excited state contamination, we do not claim to have a rigorous determination of the spectrum, since the creation and annihilation operators we have used are purely local three-quark operators which are known to have poor overlap with the nucleon-pion scattering states~\cite{Lang:2012db}.
However, since only the transition excited states depend upon the current insertion time $\t$, we can confidently separate the excited states into these various classes of excited state contamination.

\begin{figure*}
\begin{tabular}{cc}
\includegraphics[width=\columnwidth]{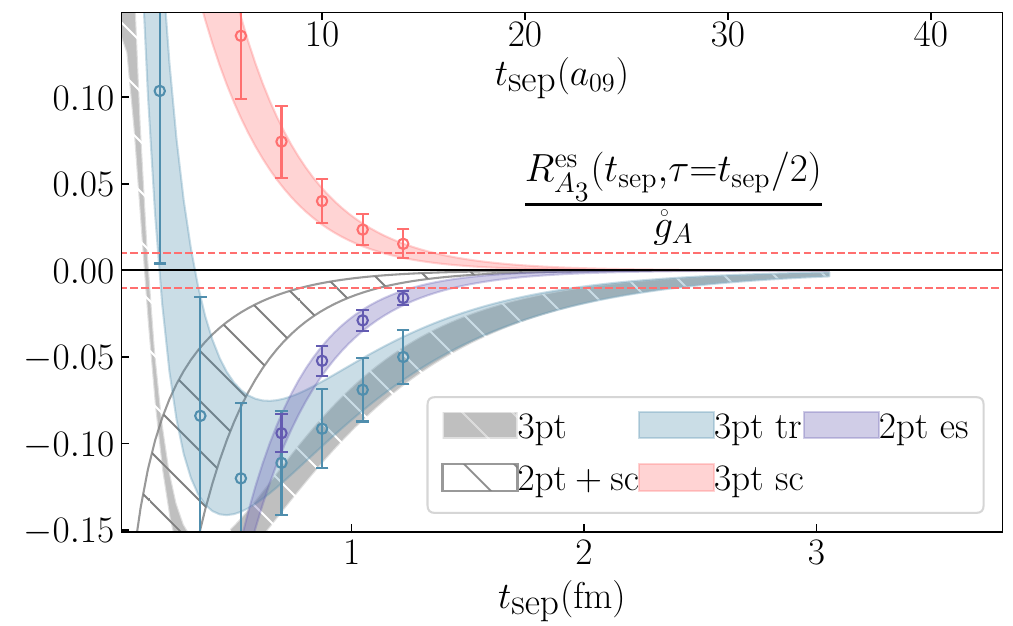}
&\includegraphics[width=\columnwidth]{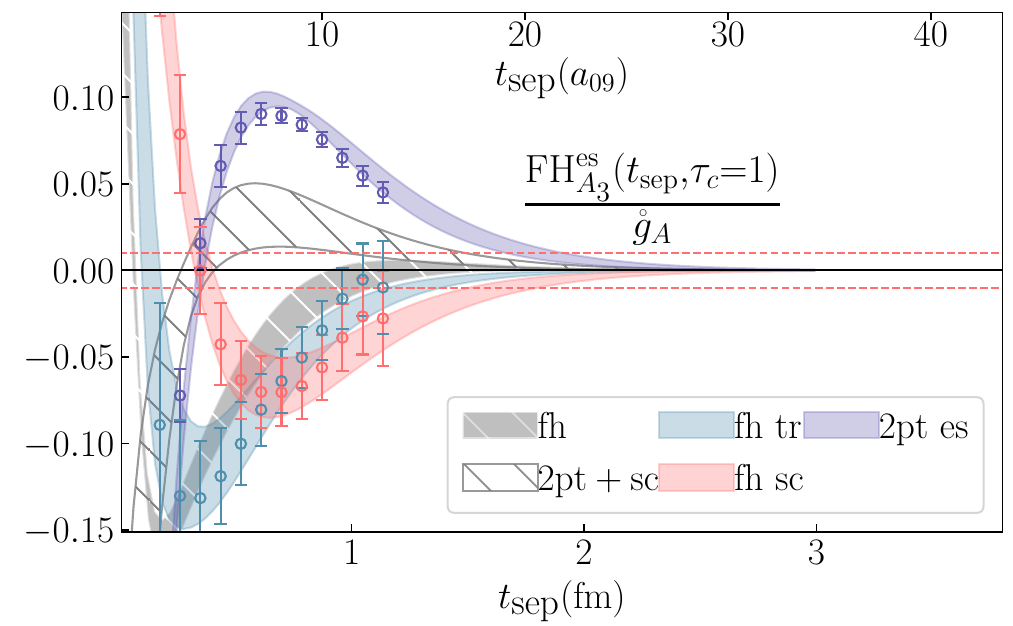}
\\
(a)& (b)
\end{tabular}
\caption{\label{fig:excited_states}
Ratio of excited state contributions to the ground state, $g_A$, for $R_{A_3}(\tsep,\t)$ (left, 3pt) and ${\rm FH}_{A_3}(\tsep,\t_c=1)$ (right, fh).
Each panel contains three types of data points: the $n$-to-$n$ scattering (sc) excited states, $C_\G^{\rm sc}/(g_A C_2)$,
the $n$-to-$m$ transition (tr) excited states, $C_\G^{\rm tr}/ (g_A C_2)$,
and the excited states from the two-point (2pt es) function, $-C_2^{\rm es}/C_2$ (and the corresponding Feynman-Hellmann versions of these ratios in the right figure).
The decomposition of the numerical data into these three classes of excited state contamination depends upon our posterior reconstruction of the correlation functions (2pt, 3pt and FH), see \eqnref{eq:excited_states}.
Because we are looking down the middle of $R_{A_3}$ (along $\tau=\tsep/2$), we only show data at even $\tsep$s in the left panel.
In addition to plotting the corresponding posterior contribution from each class of excited state, we plot two hatched bands.
The white hatched band is the sum of the 3pt sc and 2pt es contributions, which are observed to largely cancel, as indicated in \eqnref{eq:ratio_es_separated}.
The gray hatched band is the sum of all excited state contributions (the full posterior distribution, minus the ground state, normalized by the ground state).
The 3pt sc contribution is positive for $R_{A_3}(\tsep,\t)$ and it becomes negative for ${\rm FH}_{A_3}(\tsep,\t_c=1)$ at $\tsep>0.5$~fm.
The opposite is true for the 2pt es contribution.
The sum of the scattering and two-point excited state contributions is opposite in sign to the transition excited state contributions, unlike for $R_{A_3}(\tsep,\t)$ in which they are the same sign, leading to an even stronger suppression of the total excited state contributions beyond the expected suppression, as described in the text.
The horizontal red dashed lines represent the threshold of a 1\% contribution of excited state contamination to the ground state value.
}
\end{figure*}

In \figref{fig:excited_states}, we plot the percent contamination of various sources of excited state contamination to the ground state for $R_{A_3}(\tsep,\t=\tsep/2)/\mathring{g}_A$ (left) and FH$_{A_3}(\tsep,\t_c=1)/\mathring{g}_A$ (right).
The legend keys for $R_{A_3}^{\rm es}(\tsep,\t=\tsep/2)/\mathring{g}_A$ correspond to
\onecolumngrid
\begin{align}\label{eq:excited_states}
\textrm{2pt es:}\ & \phantom{\mathring{g}_A C_2(\tsep)}
    -\frac{C_2^{\rm es}(\tsep)}{C_2(\tsep)}
    &=\quad &
        \frac{C_{A_3}^{\rm data}(\tsep,\t=\tsep/2)}{\hat{\mathring{g}}_A C_2^{\rm data}(\tsep)}
        - \frac{\hat{C}_{A_3}^{\rm sc}(\tsep,\t=\tsep/2)}{\hat{\mathring{g}}_A \hat{C}_2(\tsep)}
        - \frac{\hat{C}_{A_3}^{\rm tr}(\tsep,\t=\tsep/2)}{\hat{\mathring{g}}_A \hat{C}_2(\tsep)}\, ,
\nonumber\\
\textrm{3pt sc:}\ &
    \phantom{-}\frac{C_{A_3}^{\rm sc}(\tsep,\t=\tsep/2)}{\mathring{g}_A C_2(\tsep)}
    &=\quad & \frac{C_{A_3}^{\rm data}(\tsep,\t=\tsep/2)}{\hat{\mathring{g}}_A C_2^{\rm data}(\tsep)}
        + \frac{\hat{C}_2^{\rm es}(\tsep)}{\hat{C}_2(\tsep)}
        - \frac{\hat{C}_{A_3}^{\rm tr}(\tsep,\t=\tsep/2)}{\hat{\mathring{g}}_A \hat{C}_2(\tsep)}\, ,
\nonumber\\
\textrm{3pt tr:}\ &
    \phantom{-}\frac{C_{A_3}^{\rm tr}(\tsep,\t=\tsep/2)}{\mathring{g}_A C_2(\tsep)}
    &=\quad & \frac{C_{A_3}^{\rm data}(\tsep,\t=\tsep/2)}{\hat{\mathring{g}}_A C_2^{\rm data}(\tsep)}
        + \frac{\hat{C}_2^{\rm es}(\tsep)}{\hat{C}_2(\tsep)}
        - \frac{\hat{C}_{A_3}^{\rm sc}(\tsep,\t=\tsep/2)}{\hat{\mathring{g}}_A \hat{C}_2(\tsep)}\, ,
\end{align}
\twocolumngrid
\noindent
where the $\ \hat{}\ $ denotes the posterior reconstruction of a given contribution to the correlation function (or $\mathring{g}_A$) and the ``data'' superscript denotes the numerical data.  The full correlation between the numerical data and posteriors are used in these reconstruction.

The 2pt + sc is the sum of the first two terms and the 3pt is the sum of all excited states.
The FH$_{A_3}(\tsep,\t_c=1)/\mathring{g}_A$ keys correspond to the same breakdown of various excited state contributions after they are passed through \eqnref{eq:fh_corr}.
From these plots, there are several observations and conclusions one can make:
\begin{enumerate}[leftmargin=*]
\item There is a significant cancellation between the scattering excited states and two-point excited states;

\item The transition excited states are the dominant source of excited states and they are relatively suppressed in ${\rm FH}_{A_3}(\tsep,\t_c)$ as compared to $R_{A_3}(\tsep,\t=\tsep/2)$;

\item The total excited state contamination at $\tsep\approx1$~fm is $\approx8\%$ for $R_{A_3}(\tsep,\t=\tsep/2)$ and $\approx1\%$ for ${\rm FH}_{A_3}(\tsep,\t_c)$.

\end{enumerate}

In \eqnref{eq:ratio_es_separated} we see, from the proportionality of the $n^{\rm th}$ excited state to $g_{nn}^\G - g_{00}^\G$, an explicit cancellation between the scattering (3pt sc) and two-point excited state (2pt es) contamination.
For this cancellation to be significant, as observed in the posterior determination of the two classes of excited states, one explanation is that $g_{nn}^\G\approx g_{00}^\G$ for all $n$.
For the vector current, we know that in infinite volume, the vector operator measures the iso-vector charge of the system, which in the isospin limit, is exactly equal to $g_{00}^{V_4} = 1$ (up to renormalization).
It has further been demonstrated that in the forward scattering limit, the finite volume interaction amplitudes conspire with the finite volume Lellouch-L\"{u}scher factors to ensure that $g_{nn}^{V_4}=g_{00}^{V_4}$ for two-particle states~\cite{Briceno:2019nns}.

This result, while not surprising, demonstrates charge conservation in finite volume by allowing one to relate $n$-point Greens functions to $(n-1)$-point Greens functions through the Ward-Takahashi equations.
In the case of the axial-vector matrix element, it is plausible that in the forward scattering limit, the corrections to such a relation arising from the partially-conserved nature of the axial-vector current would also vanish, such that in this limit, the Lellouch-L\"{u}scher factors are similarly removed.
For example, for the nucleon, we know the induced pseudo-scalar form factor, which would provide a correction to the Ward-Takahashi equations, vanishes in the forward limit.
We leave a detailed study of this question to future work.

In the left panel of \figref{fig:excited_states} we see that the scattering excited state contribution (the upper (red) band/data) and the contribution from the excited states coming from the two-point function (the middle (purple) band/data) are roughly equal and opposite in sign.
The hatched curve, which is mostly white with gray hash lines, is the sum of these two contributions and it lies between them.
We observe that the sum of these two sources of excited state contributions decay to a 1\% correction (the horizontal dashed (red) lines) at $\tsep\approx 1$~fm.
In contrast, the transition (3pt tr) excited states depicted by the lowest (blue) band/data do not decay to the 1\% level until $\tsep\approx2.2$~fm.

In the right panel of \figref{fig:excited_states} we show the same excited state breakdown for the Feynman-Hellmann correlation function.
The sign of the scattering and two-point excited state contributions both change compared to $R_{A_3}(\tsep,\t=\tsep/2)$.
This sign change can be understood from \eqnref{eq:fh_corr}.
It is also interesting to note that the magnitude of the transition excited state contributions at $\tsep\approx1$~fm go from $\approx7.5\%$ for $R_{A_3}(\tsep,\t=\tsep/2)$ to $\approx2.5\%$ for FH$_{A_3}(\tsep,\t_c=1)$.
Finally, for $R_{A_3}(\tsep,\t=\tsep/2)$, the sum of the scattering and two-point excited state contributions is the same sign as the sum of transition excited states, while for FH$_{A_3}(\tsep,\t_c=1)$ it is the opposite sign at intermediate and large $\tsep$.
Thus, the total excited state contamination decays to the 1\% level for the Feynman-Hellmann correlation function at $\tsep\approx1$~fm while this does not happen until $\tsep>2$~fm for $R_{A_3}(\tsep,\t=\tsep/2)$.

There is no proof that this fortunate cancellation must happen.
However, an examination of our results in Ref.~\cite{Chang:2018uxx} shows a consistent picture that the excited state contamination of the FH$_{A_3}(\tsep,\t_c={\rm none})$ correlation function decays to the 1\% level by roughly 1 fm over a broad range of pion masses, $130\lesssim m_\pi \lesssim 400$~MeV.

\subsection{\label{sec:latetime_results} Comparison with late-time only results}
In this section, we perform the correlator analysis on late time separation data, $\tsep = [10, 12, 14]$, which is directly comparable to the results in Refs.~\cite{Bhattacharya:2016zcn,Gupta:2018qil} that were performed on the same ensembles, and otherwise mimicking the more common strategy used for example in Refs.~\cite{Capitani:2012gj,Horsley:2013ayv,Bali:2014nma,Abdel-Rehim:2015owa,Alexandrou:2017hac,Capitani:2017qpc,Park:2021ypf}.

The results are obtained with a simultaneous fit to the two- and three-point functions using the same priors as our main fit, given in \appref{sec:analysis:priorselection}.
We also demonstrate the sensitivity of the posterior distributions of the ground state parameters to changes in the input priors as in \appref{sec:analysis:stability}.
Because we have only three values of $\tsep$, we use a minimal number of excited states: with one excited state, there is only one degree of freedom in describing the $\tsep$ dependence of $R_\G(\tsep,\t)$, after which, the analysis relies upon the priors for the excited state matrix elements.

\begin{figure}
\includegraphics[width=\columnwidth]{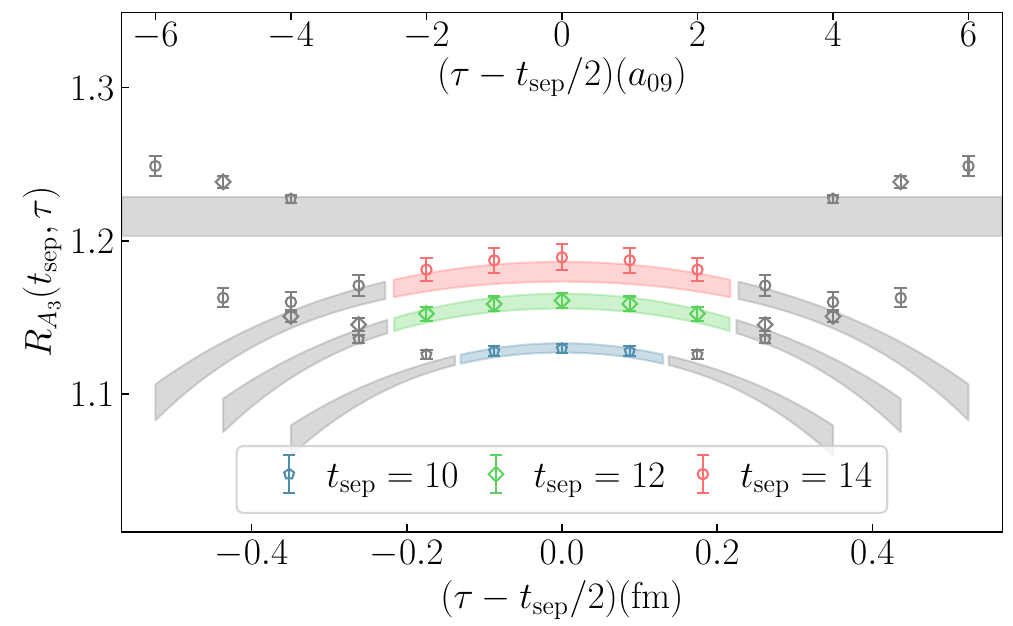}
\caption{\label{fig:fit_region_late_tsep}
We plot the numerical data for $R_{A_3}(\tsep,\t)$ and resulting posterior fit bands obtained from a combined two-point and three-point analysis when only $\tsep=[10,12,14]$ are used in the analysis.  The inner (colored) numerical data are used in the analysis.  There is a break in the fit band and the outer (gray) band and data indicate regions of $\t$ not used in the analysis.
The horizontal gray band is the resulting posterior value of the ground state, $\mathring{g}_A$.
}
\end{figure}
\begin{figure}
\includegraphics[width=\columnwidth]{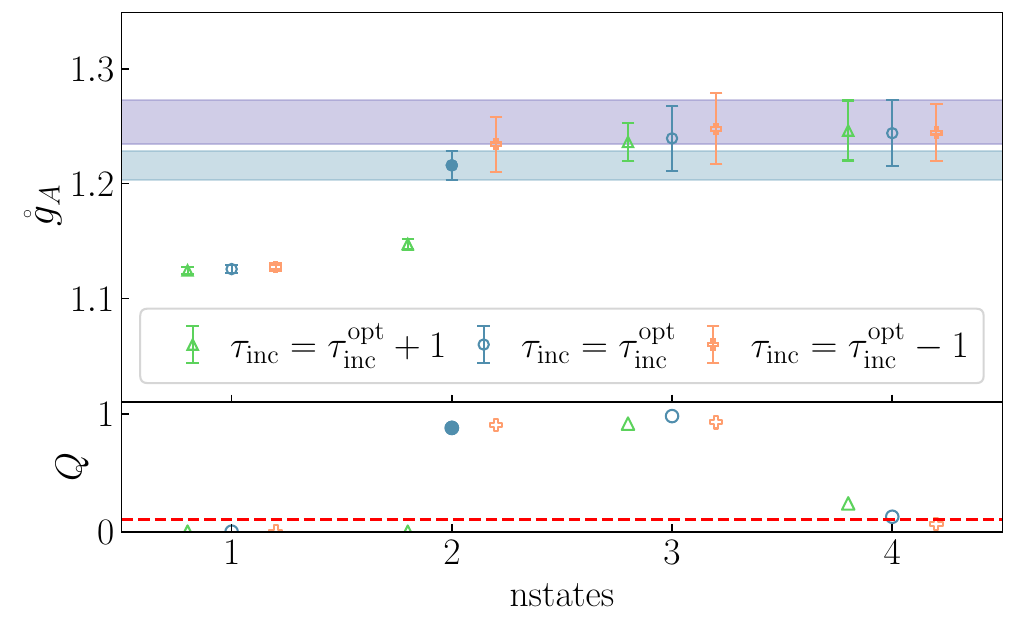}
\caption{\label{fig:fit_late_tsep_comparison}
The stability of $\mathring{g}_A$ versus the number of states and current insertion times included in the analysis. $\tinc^{\rm opt}$ denotes our optimal choice of included data.
The bottom panel shows the corresponding fit quality $Q$.
The lower horizontal band indicates the optimal fit from the late-$\tsep$ only analysis while the upper horizontal band above it represent the optimal fit from the full analysis presented in \secref{sec:full_results}.
}
\end{figure}

With the quark smearing we have used we find that we must apply a relatively aggressive truncation on the current insertion time in order for the model to describe the numerical data, restricting the analysis to the ``center'' of the current insertion time $\t$.
We illustrate the region of $\t$ used for each $\tsep$ in \figref{fig:fit_region_late_tsep} and depict the stability of the ground state axial matrix element in \figref{fig:fit_late_tsep_comparison}.
The optimal result is chosen as the fit with the largest amount of numerical data while maintaining a good fit quality (cfr. the lower panel of \figref{fig:fit_late_tsep_comparison}).
We refer to the optimal choice of data included for this 2-state model fit as $\tinc=\tinc^{\rm opt}$, but we also show results for the ground state posteriors when one more ($\tinc^{\rm opt}+1$) or one less ($\tinc^{\rm opt}-1$) value of $\t$ is included in the fit.
While not depicted, the ground state posteriors are also stable under variation of the choice of $\tsep^{\rm min}$ in the two-point function.

We observe that for this specific dataset, the late-$\tsep$ result is in $\approx2\s$ tension with the result from the full analysis that includes more values of $\tsep$, and is clearly stable over a variety of data truncations and model variations.
This is indicative that late-$\tsep$ data can be subject to correlated fluctuations which are difficult to identify without having results at many values of $\tsep$.

\begin{figure}
\includegraphics[width=\columnwidth]{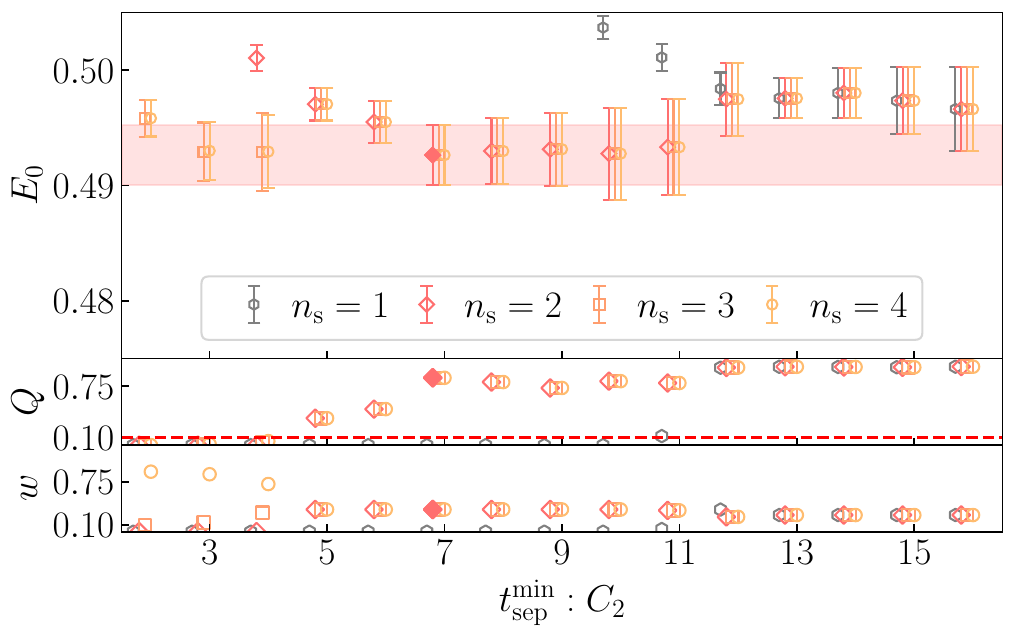}
\caption{\label{fig:fit_late_tsep_2pt}
The extracted ground state energy as a function of $\tsep^{\rm min}$ and the number of states ($n_s$) used in the analysis.
The late-$\tsep$ data is subject to a correlated fluctuation, as observed in the $2\s$ increase in $E_0$ around $\tsep^{\rm min}=12$.
}
\end{figure}

The same systematic effect can be seen in a simpler case using a fit to the two-point correlation function.
In \figref{fig:fit_late_tsep_2pt}, we show the result of the ground state energy extracted from the two-point correlation function under varying $\tsep^{\rm min}$.
We highlight the best fit ground-state energy, which is supported by a robust plateau in the dimension of $\tsep^{\rm min}$ with high $Q$-value.
However, we see that at late time beyond 1~fm, a second plateau develops approximately two standard deviations above our best fit. This second plateau cannot be physical and simply reached later in time due to latent excited state contamination, as the 2-pt correlator is positive definite and therefore required to approach the ground state from above.
This second ``stable'' plateau is also the logical result for a one-state fit.
Analogous to the three-point analysis, if the model fails to describe excited-state contaminations, the model must extract ground-state parameters at late time.
These values are then sensitive to uncontrolled statistical fluctuations.
Also note that we observe the fluctuation in the effective mass occurs approximately 0.2~fm later than the matrix elements, see the top panel of \figref{fig:effective} in \appref{app:analysis_details}.
Similar to the three-point analysis, in the absence of a more holistic view as granted by analyzing more source-sink separation data, there is no measure on the size of how large the potential underestimation of errors are.

Without using more than three values of $\tsep$, it is not possible to identify if one is susceptible to such a fluctuation, and therefore, it is not possible to fully quantify the uncertainty on the posterior distribution of the ground state parameters.

\subsection{Comparison with the 2-state FH analysis}
In this final section, we compare our result with that from Ref.~\cite{Chang:2018uxx} on the same a09m310 ensemble.
Those results were obtained using the Feynman-Hellmann method as described in Ref.~\cite{Bouchard:2016heu}, which is the same as \eqnref{eq:fh_corr} with the sum over current insertion time running over the full time extent, including contributions from contact operators at $\t = [0,\tsep]$ and from out-of-time region, $\t<0$ and/or $\t>\tsep$.
We applied a two-state model and a frequentist analysis on the data from the two-point and Feynman-Hellmann correlation functions.

\begin{figure}
\includegraphics[width=\columnwidth]{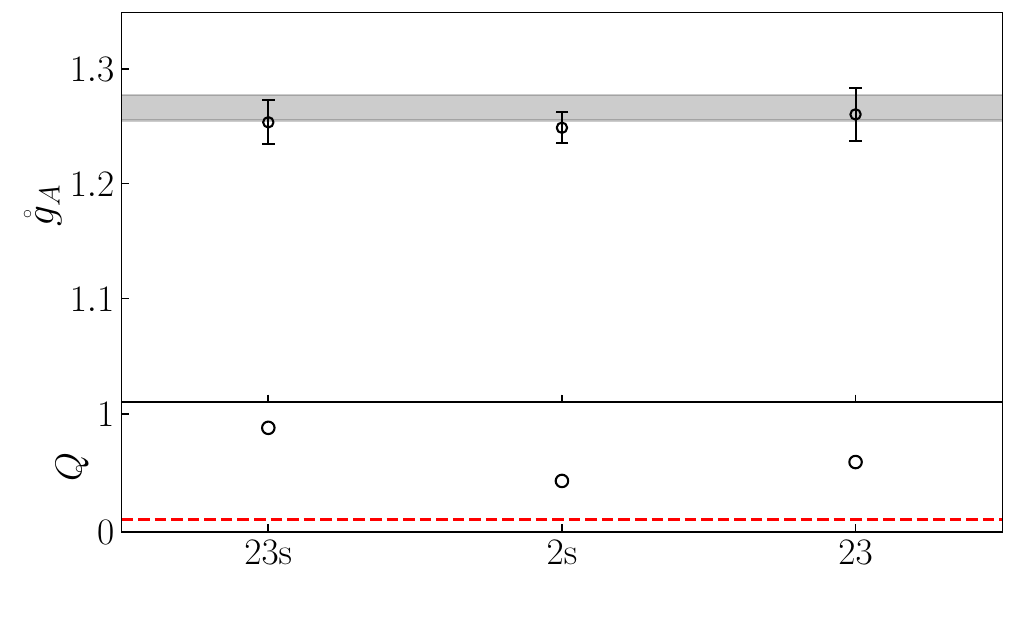}
\caption{\label{fig:comparison}
Comparison of $\mathring{g}_A$ determined from the three analysis strategies described in the text with our result from Ref.~\cite{Chang:2018uxx} which utilized the Feynman-Hellmann correlator with a sum over all current insertion times~\cite{Bouchard:2016heu} (note that the quark smearing in the present analysis is different from that used in Ref.~\cite{Chang:2018uxx}).
}
\end{figure}

In \figref{fig:comparison} we show the extracted value of $\mathring{g}_A$ with three different fully-correlated analysis strategies:
\begin{enumerate}
\item[\textbf{23s}]: Fit to $C_2(\tsep)$, $R_\G(\tsep,\t)$ and FH$_\G(\tsep,\t_c=1)$;

\item[\textbf{2s}]: Fit to $C_2(\tsep)$ and FH$_\G(\tsep,\t_c=1)$;

\item[\textbf{23}]: Fit to $C_2(\tsep)$ and $R_\G(\tsep,\t)$.
\end{enumerate}
The horizontal gray band is the result from Ref.~\cite{Chang:2018uxx}.
These results show the consistency between the four computational and analysis strategies.
Even though the FH$_\G(\tsep,\t_c)$ correlation functions are constructed from the $R_\G(\tsep,\t)$ correlation functions, they are subject to a different pattern of excited state contamination (cfr. \secref{sec:spectral_decomp}).
Therefore it is useful to include them in the fully-correlated analysis.
The combined fit yields a robust and consistent fitting strategy and precise extractions of hadronic matrix elements at zero-momentum transfer.

\begin{figure}
\includegraphics[width=\columnwidth]{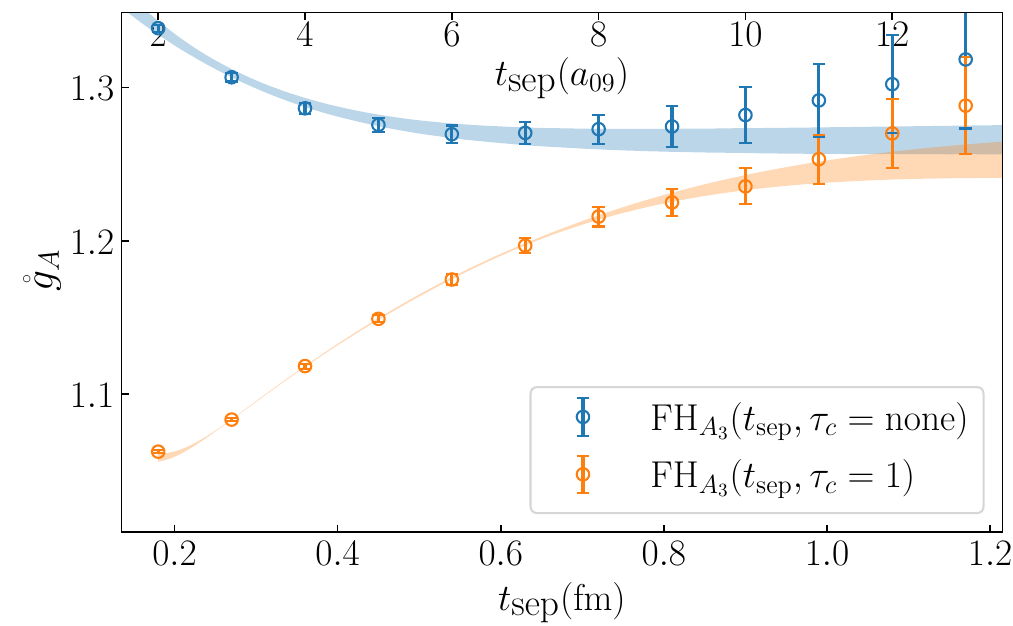}
\caption{\label{fig:fit_comparison}
Comparison of our FH$_{A_3}(\tsep,\t_c={\rm none})$ numerical results and analysis from Ref.~\cite{Chang:2018uxx} using a two-state frequentist analysis (top data/band) to the FH$_{A_3}(\tsep,\t_c=1)$ numerical data and posterior description of the correlation function from the fully-correlated five-state Bayesian analysis to $C_2(\tsep)$, FH$_{\G}(\tsep,\t_c=1)$ and $R_\G(\tsep,\t)$ in the present work (bottom data/band).
}
\end{figure}

In \figref{fig:fit_comparison}, we present the posterior correlation function on top of the numerical data for both the results in Ref.~\cite{Chang:2018uxx} as well as from the FH$_{A_3}(\tsep,\t_c=1)$ dataset in the present work.
Using the bootstrap results, the correlated ratio of $g_A$ determined previously and now is
\begin{equation}
\frac{g_A\text{\cite{Chang:2018uxx}}}{g_A[\rm present~work]} = 1.009(31)\, ,
\end{equation}
showing the statistical consistency of the ground state matrix element extracted from these two different methods on the same ensemble.

The FH$_{A_3}(\tsep,\t_c={\rm none})$ dataset from Ref.~\cite{Chang:2018uxx} is observed to have significantly less excited state contamination at early $\tsep$ as compared to FH$_{A_3}(\tsep,\t_c=1)$.
This is what enabled a two-state analysis in Ref.~\cite{Chang:2018uxx}.
The combined fit (23s) in the present work enables a determination of significantly more excited state parameters through the precise early-$\tsep$ data.
Moreover, as shown in this section, a confident extraction of the ground state posterior distributions is attained: as depicted in the figure, both strategies yield precise and consistent extractions of the large-$\tsep$ extrapolation of the results.
The former result is more economical for obtaining a precise value of $\mathring{g}_A$, while the method in this work also enables a determination of non-zero momentum transfer results which can be used to determine the form factors.

\section{\label{sec:conclusions} Observations and Conclusions}
In this work, we have computed the three-point correlation functions that are used to determine the nucleon axial charge for 13 values of the source-sink separation time in the range $\tsep\approx0.17 - 1.22$~fm on an ensemble with $a\approx0.09$~fm and $m_\pi\approx310$~MeV.
This large numerical dataset (\figref{fig:psychedelic_moose}) has enabled us to robustly determine the ground state mass and matrix elements with a fully-quantified systematic uncertainty arising from excited states.
Important to this study is the use of a fully-correlated global minimization of the two-point and respective three-point correlations functions.

We were further able
to quantify the excited state contribution to the correlation function with greater detail than has been previously achieved in the literature.
We were able to demonstrate that the ground state parameters are stable against the model of excited states (\eqnref{eq:spectrum_parameterization} and \figref{fig:excited_state_model}), the number of excited states used in the analysis, and truncations of either small $\tsep$ or large $\tsep$ results as well as through the use of only even or odd values of $\tsep$ results (\figref{fig:stability}).

A re-writing of the spectral decomposition of the three-point correlation functions revealed a prospective cancellation between two classes of excited states arising from the $n$-to-$n$ scattering states and those arising from the two-point correlation function used to construct $R_\G(\tsep,\t)$, \eqnref{eq:ratio_es_separated} (there is an exact cancellation for the vector matrix element for which $g_{nn}^{V_4} = g_{00}^{V_4} = 1$ up to renormalization).
For our calculation, this cancellation seems to materialize for $g_A$ (\figref{fig:excited_states}) leaving the dominant excited state contributions to be the $n$-to-$m$ transition terms.
In the standard fixed source-sink separation method, the excited state contributions do not decay down to the 1\% contamination level until $\tsep\gtrsim2$~fm while in the Feynman-Hellmann correlation function, \eqnref{eq:fh_corr}, they decay to the 1\% level at $\tsep\approx1$~fm.
We further observed that the Feynman-Hellmann correlation function constructed with a sum over the entire time extent of the lattice~\cite{Bouchard:2016heu} (FH$_\G(\tsep,\t_c={\rm none})$), rather than just between the source and the sink (FH$_\G(\tsep,\t_c=1)$), leads to an even further suppression of excited state contamination (\figref{fig:fit_comparison}).
This stronger cancellation of excited states enabled us to compute $g_A$ with a $\approx1\%$ uncertainty by utilizing early-time data~\cite{Berkowitz:2017gql,Chang:2018uxx,Berkowitz:2018gqe,Walker-Loud:2019cif}, which demonstrated this higher suppression of excited states over a broad range of pion masses, $130\lesssim m_\pi \lesssim 400$~MeV.

By utilizing a large number of early to mid-time data, one is able to detect if the late-time data is subject to a correlated fluctuation which might otherwise bias the determination of the ground state matrix elements (Figs.~\ref{fig:fit_late_tsep_comparison} and \ref{fig:fit_late_tsep_2pt}).
With only three values of $\tsep$ at late time, it is not possible to perform such a data truncation study that could identify this issue.

We have found that including the Feynman-Hellmann correlation functions in a global fully correlated analysis, even though they are constructed from the three-point functions as in \eqnref{eq:fh_corr} and are highly correlated with $R_\G(\tsep,\t)$, improves the stability and precision of the extracted ground state parameters (\figref{fig:comparison}).  This improvement arises because the excited states present themselves differently in the two sets of matrix-element correlation functions.

Even though the Feynman-Hellmann correlation function has greater noise than the standard three-point correlator at equal $\tsep$ and statistics (compare, for example, $\tsep=1$~fm. in \figref{fig:psychedelic_moose}), the more rapid decay of excited states demonstrated in \secref{sec:es_breakdown} means that this strategy mitigates the exponential degradation of the signal-to-noise of the nucleon's two- and three-point correlation functions as $\tsep$ is increased, which requires exponentially more computational resources to control the stochastic precision.  The strategy we present in this work offers a more economical method of obtaining the ground state matrix elements than that which is more commonly advocated for in the literature, which is to use high-statistics calculations at $\tsep\approx1-2$~fm or larger.
In future work, we will investigate the same strategy for non-zero momentum transfer correlation functions which are used to determine the nucleon form factors.

\bigskip\noindent\textbf{Data availability:}
The computations were performed utilizing \texttt{LaLiBe}~\cite{lalibe} which utilizes the \texttt{Chroma} software suite~\cite{Edwards:2004sx} with \texttt{QUDA} solvers~\cite{Clark:2009wm,Babich:2011np} and HDF5~\cite{hdf5} for I/O~\cite{Kurth:2015mqa}.
They were efficiently managed with \texttt{METAQ}~\cite{Berkowitz:2017vcp,Berkowitz:2017xna} and status of tasks logged with EspressoDB~\cite{Chang:2019khk}.
The final extrapolation analysis utilized \texttt{gvar}~\cite{gvar} and \texttt{lsqfit}~\cite{lsqfit}.
The analysis and data for this work can be found at \url{https://github.com/callat-qcd/project_fh_vs_3pt}.

\acknowledgments
We thank O.~B\"ar, R.~Brice\~no, R.~Gupta, M.~Hansen and A.~Jackura for enlightening conversations and correspondence.
We thank B.~H\"{o}rz for help cross-checking the three-point code in LaLiBe and generating some of the results.
We thank the MILC Collaboration for use of their HISQ gauge ensembles.

Computing time for this work was provided through the Innovative and Novel Computational Impact on Theory and Experiment (INCITE) program and the LLNL Multiprogrammatic and Institutional Computing program for Grand Challenge allocations on the LLNL supercomputers.  This research utilized the  NVIDIA GPU-accelerated Summit supercomputer at Oak Ridge Leadership Computing Facility at the Oak Ridge National Laboratory, which is supported by the Office of Science of the U.S. Department of Energy under Contract No. DE-AC05-00OR22725 as well as the Lassen supercomputer at Lawrence Livermore National Laboratory, which is operated by the National Nuclear Security Administration of the U.S. Department of Energy under Contract No. DE-AC52-07NA27344.

This work was supported in part by the Berkeley Physics International Education Program (JH),
the Berkeley Laboratory Undergraduate Research Program (IC);
the NVIDIA Corporation (MAC),
the Alexander von Humboldt Foundation through a Feodor Lynen Research Fellowship (CK),
the U.S. Department of Energy, Office of Science, Office of Nuclear Physics under Award Numbers
DE-AC02-05CH11231 (CCC, CK, AWL),
DE-AC52-07NA27344 (DAB, DH, ASG, PV),
DE-FG02-93ER-40762 (EB),
DE-AC05-06OR23177 (CJM);
DE-SC00046548 (ASM);
the DOE Early Career Award Program (AWL),
and the U.K. Science and Technology Facilities Council grants ST/S005781/1 and ST/T000945/1 (CB);

\appendix

\section{Discrete symmetry systematics\label{app:discrete_symmetry}}
\noindent\textit{``Coherent sink technique''}~\cite{Bratt:2010jn}:
We reduce the numerical cost of the computations by solving for a single sequential-propagator from many sequential sinks simultaneously.
We found that we can combine eight sinks into a single coherent sink generated with two sources per timeslice, with a 10-20\% loss in statistical precision as compared to solving a single sequential-propagator for each of the eight sources separately.
For each $t_0$, a random origin (O) is chosen and then the antipode (A) location is also chosen~\cite{Berkowitz:2019yrf}
\begin{align*}
s_{\rm O}(t_0) &= (x_0, y_0, z_0)\\
s_{\rm A}(t_0) &= \left[ s_{\rm O}(t_0) + \frac{L}{2}(1,1,1) \right]\ {\rm mod}\ L\, .
\end{align*}
We repeat this for four values of $t_0$ spaced by $T/4$.
We generate all 16 sources by running a second set of eight sources shifted by $T/8$ from the first set of sources.

\textit{Spin-averaging:}
We find that combining the spin-up--to--spin-up and spin-down--to--spin-down correlation functions (with a $+$ sign for $V_4$ and a $-$ sign for $A_3$) leads to a near perfect $\sqrt{2}$ reduction in the stochasitc uncertainty of the numerical data.
We further observe that the non-symmetric behavior of $R_\G(\tsep,\t)$ about $\t=\tsep/2$ for larger values of $\tsep$, which must vanish in the infinite statistics limit, is less pronounced when we perform this spin-averaging.

\textit{Time-reversal symmetry:}
We find that combining the backwards temporal propagation of the negative-parity two- and three-point functions -- the time-reversed correlation functions (negative parity three-point functions with negative values of the source-sink separation time $\tsep = t_{\rm snk} - t_{\rm src} < 0$) with the positive parity three-point functions generated with $\tsep>0$ leads to a near-perfect $\sqrt{2}$ reduction in stochastic uncertainty, allowing us to make use of both the positive and negative parity components of the quark propagators.

\section{Analysis Details\label{app:analysis_details}}
In this appendix, we discuss in detail the analysis of the various correlation functions and what led us to our final strategy presented in \secref{sec:lattice_calc}.

To analyze the two-point (2pt), \eqnref{eq:2pt}, three-point (3pt), \eqnref{eq:ratio} and Feynman-Hellmann (FH), \eqnref{eq:fh_corr} correlation functions and determine the parameters of the fit model, we perform a maximum-likelihood Bayesian analysis.
We explore analyzing three combinations of correlation functions in a global (simultaneous) analysis:
\begin{enumerate}
\item[\textbf{23s}]: $C_2(\tsep)$, $R_\G(\tsep,\t)$ and FH$_\G(\tsep,\t_c=1)$;

\item[\textbf{2s}]: $C_2(\tsep)$ and FH$_\G(\tsep,\t_c=1)$;

\item[\textbf{23}]: $C_2(\tsep)$ and $R_\G(\tsep,\t)$.
\end{enumerate}

\subsection{Prior selection \label{sec:analysis:priorselection}}

\begin{figure}
\includegraphics[width=\columnwidth]{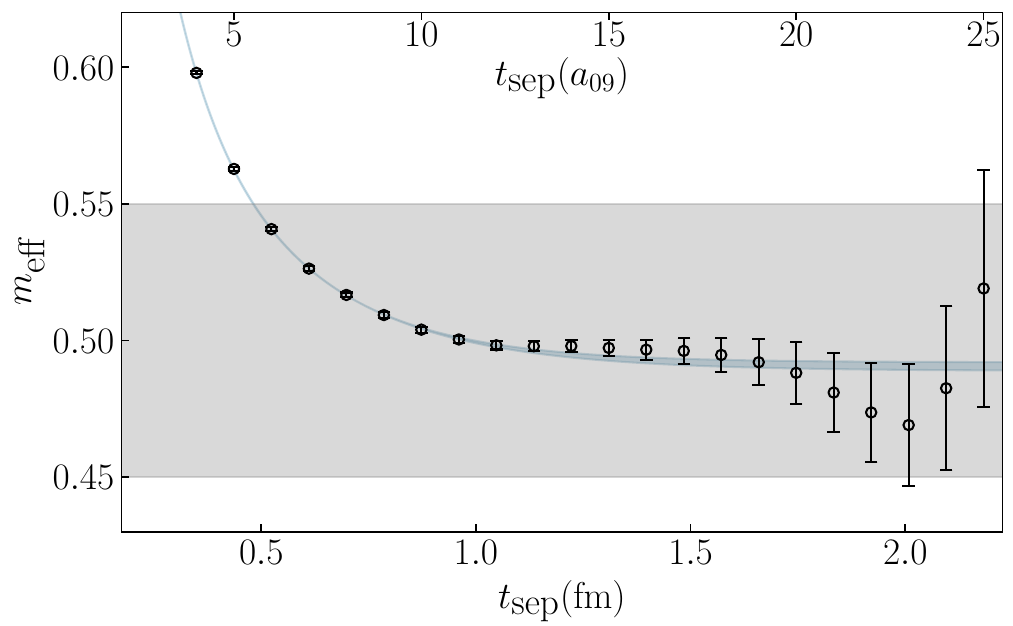}
\includegraphics[width=\columnwidth]{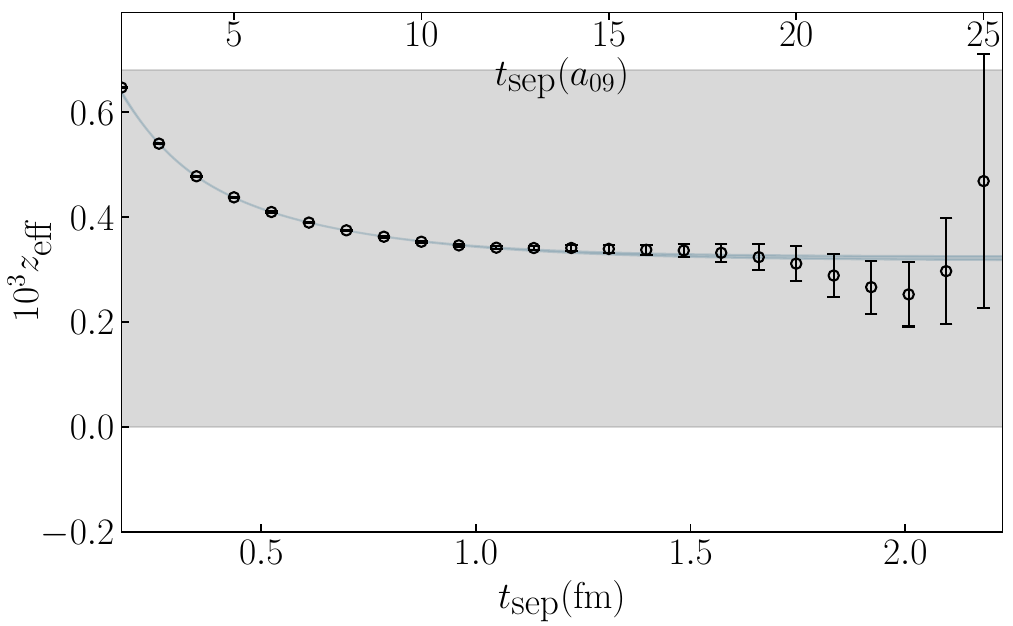}
\includegraphics[width=\columnwidth]{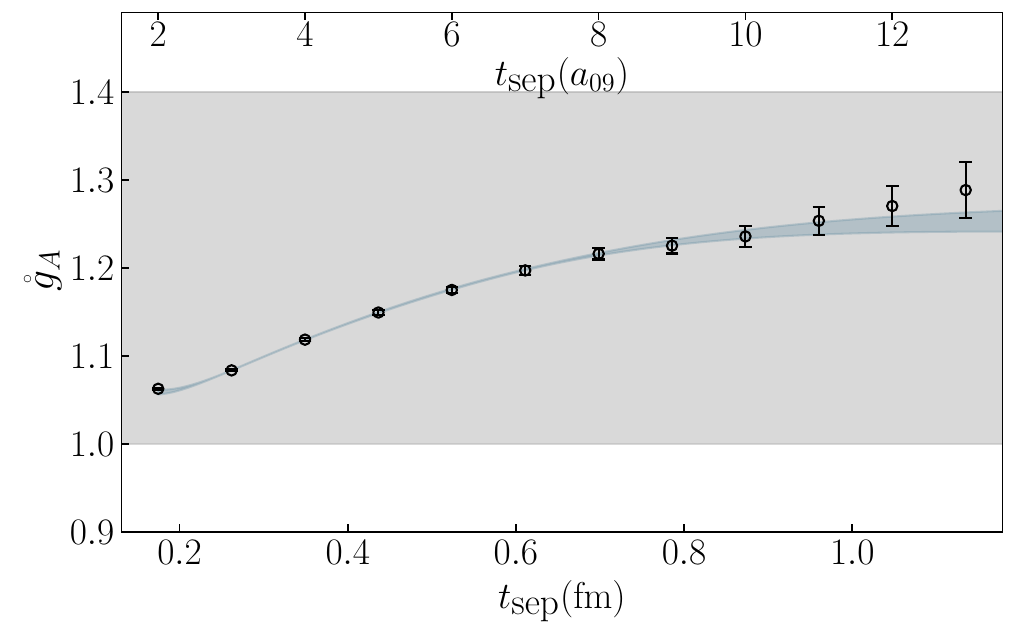}
\caption{\label{fig:effective} The effective mass (Top), overlap factor $z$ (Middle) and $g_A$ correlation functions (Bottom).
The wide horizontal gray bands are the chosen ground state priors, while the blue overlay bands are the reconstructed effective quantities using the posterior distributions from the final analysis.
}
\end{figure}

The first step in the analysis is to choose prior distributions for the parameters.
In order to estimate the ground state priors, we use the effective mass, the effective overlap and an effective $g_A$ plot:\footnote{We do not show the effective $g_V$ plot since $g_V$ is very close to 1 with our lattice action.}
\begin{align}
m_{\rm eff}(\tsep) &= \ln \left( \frac{C_2(\tsep)}{C_2(\tsep+1)} \right)\, ,
\nonumber\\
z^2_{\rm eff}(\tsep) &= e^{m_{\rm eff}(\tsep) \tsep} C_2(\tsep)\, ,
\nonumber\\
g_{A}^{\rm eff} &= {\rm FH}_{A_3}(\tsep,\t_c=1)\, .
\end{align}
In \figref{fig:effective}, we plot these effective quantities which all asymptote to their ground state values in the large $\tsep$ limit.
We choose conservative ground state priors to be
\begin{align}\label{eq:gs_priors}
&\tilde{E}_0 = 0.50(5)\, ,&
&\tilde{g}_A = 1.2(2)\, ,&
\nonumber\\
&\tilde{z}_0 = 0.00034(34)\, ,&
&\tilde{g}_V = 1.0(2)\, ,&
\end{align}
which are plotted as the wide gray horizontal bands.  We also plot the resulting posterior distribution of the effective quantities resulting from the parameters of our final analysis.

For the excited state energies we explore three models of excited states, \eqnref{eq:spectrum_parameterization}.
As shown in \figref{fig:excited_state_model}, the posterior energies are largely insensitive to the model.
We therefore focus the discussion on the spectrum of our chosen model, $\D E_n = 2m_\pi$ for all $n$.
The lowest-lying excitation is a nucleon-pion P-wave or a nucleon with two-pions at rest, up to interaction energies which are a small fraction of the total energy.
For our $m_\pi L$, these two energy levels are practically degenerate and therefore modeled as a single excitation.
We prior all the $\D E_n$ with a log-normal distribution, $\ln(\D E_n) = (\ln(2m_\pi), 0.5)$ such that the resulting energies, $E_n = E_0 + \sum_{l=1}^n \D E_l$ are ordered.

While the creation and annihilation operators are conjugate to each other, this does not fix the absolute sign of $z_n$.
Further, there is a redundancy in sign of the transition matrix elements: if all the overlap factors are taken to be positive, $z_n > 0$, a negative contribution will manifest in a negative value of $g_{nm}$.
Only the combination $z_n z^\dagger_m g_{nm}$ has a well defined sign.

To be conservative, we prior the central values of the excited state overlap factors with a central value of $0$.
For the first excited state, we choose a slightly smaller prior width with respect to the ground state and for the higher excited states we use again a slightly smaller width:
\begin{align}
&\tilde{z}_1 = 0(0.00025)\, ,&
&\tilde{z}_{n\geq2} = 0(0.00015)\, .
\end{align}
These slight reductions are motivated by the use of a smeared quark source, which suppresses excited state overlap factors compared to the ground state.

For the vector matrix elements, the conserved charge protects the charge of all states to be $g^V_{nn}=1$, even in finite volume~\cite{Briceno:2019nns}.
For the transition matrix elements, we postulate that these are the same order of magnitude, but with an unknown sign.
For the axial-vector matrix elements, we postulate the excited state matrix elements and transition matrix elements are of the same order of magnitude as $g_A$.
This leads us to the prior values of
\begin{align}
&g^V_{nn} = 1.0(2)\, ,&
&g^A_{00} = 1.2(2)\, ,&
&g^A_{nn} = 0(1) \textrm{ for } n>0\, ,
\nonumber\\
&g^V_{nm} = 0(1)\, ,&
&g^A_{nm} = 0(1)\, .&
\end{align}
A complete list of the prior and posterior distributions of all fit parameters is provided in \appref{app:priors}.

When performing a multi-exponential fit, it is expected that the highest state used in the analysis serves as a ``garbage'' can that is contaminated by the tower of more highly-excited states not included in the analysis.
The 3pt and FH correlation functions have different parametric dependence upon the excited states.
Therefore, when exploring the parameter space of fits, if the number of states used for example in the FH correlation functions differs from the 2pt function, we allow the highest lying state in each correlation function to have different priors.
Specifically, if the 2pt uses five states and the FH uses three states, then the energy and overlap priors of the third FH state are decoupled from the third state of the 2pt function.
When the number of states used is the same, we find we are able to describe the correlation functions well when keeping the highest garbage-can state the same in all correlation functions.

\subsection{Sensitivity analysis \label{sec:analysis:stability}}

For each set of correlation functions, the best fit is chosen after a careful study of the posterior distribution sensitivity on input fit parameters including fit ranges, number of states in the fit model, prior widths, and model dependence of the excited state spectrum. In the following sections, we discuss the best fit under the context of three different fit strategies, \textbf{23}, \textbf{2s} and \textbf{23s}, and then the costs and benefits of these different strategies.

\subsubsection{Fit region and $n_{\textrm{state}}$ stability analysis}

Due to the structure of excited state contamination, the posterior distributions are most sensitive to changes in $\tsep^{\rm min}$, the minimum source-sink time, and the number of excited states $n_{\textrm{states}}$ ($n_s$) included in the model.
In this section we discuss the stability of the best fit under changing these two dimensions.

\begin{center}{\textit{Two-point with Feynman-Hellmann analysis} (\textbf{2s})}\end{center}

\begin{figure}
\begin{center}
\includegraphics[width=\columnwidth]{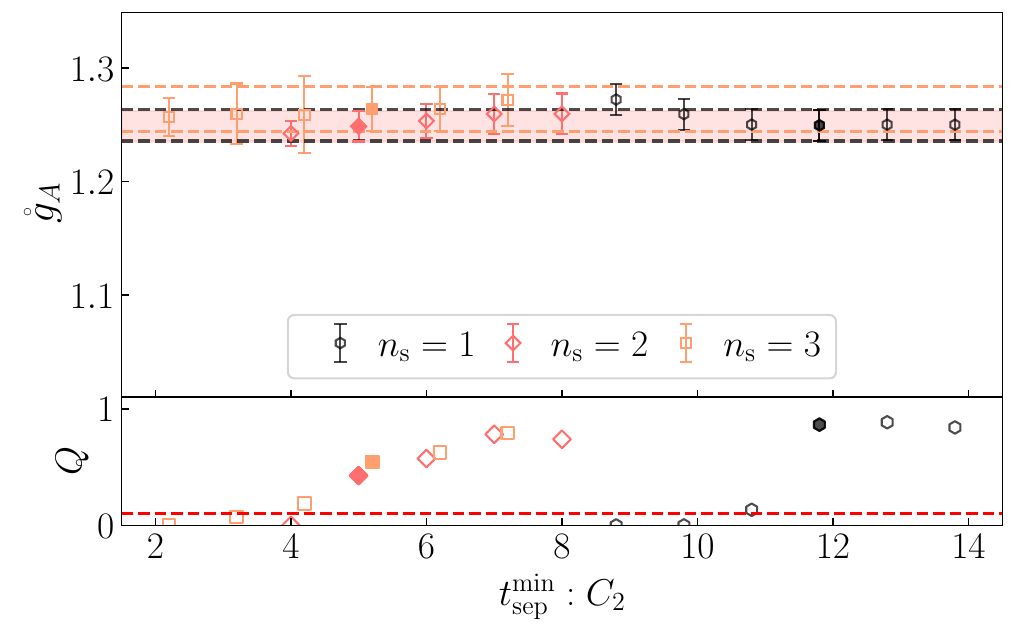}
\includegraphics[width=\columnwidth]{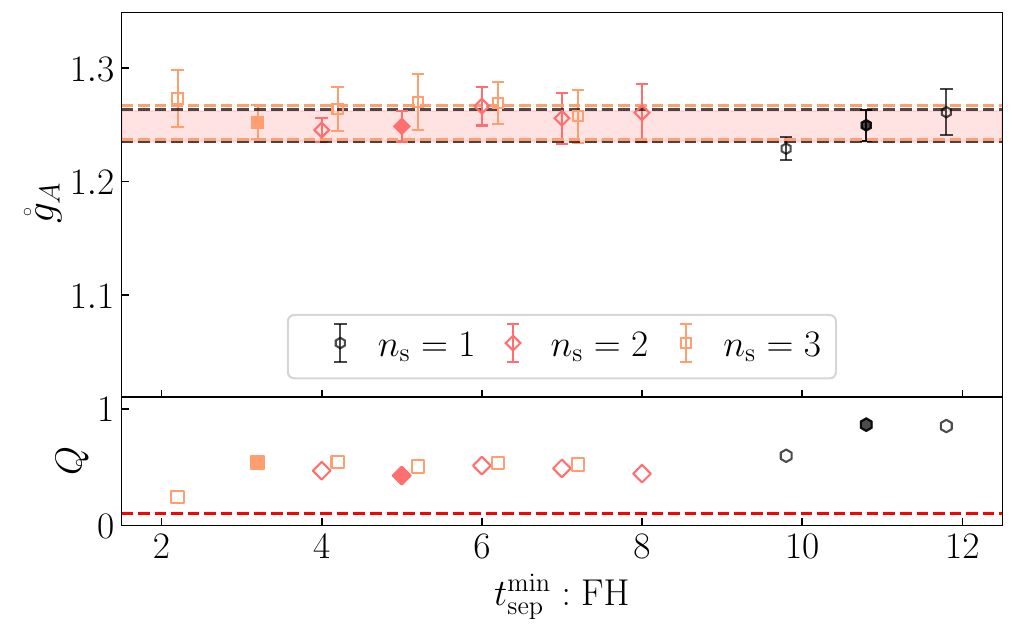}
\caption{\label{fig:tmin_nstate_stability_2s} Stability analysis for the axial coupling for the combined two-point and Feynman-Hellmann fit for (Top) varying $\tsep^{\rm min}$ of the two-point function only and (Bottom) Feynman-Hellmann correlation function only. In both cases, results of $n_s=[1, 2, 3]$ state fit ansatzes are shown. The best fit for a given $n_s$ is highlighted by a solid marker and corresponding horizontal band. The filled horizontal red band for the $n_s=2$ result highlights the best fit out of the entire explored parameter space. The corresponding $Q$-values are also provided.}
\end{center}
\end{figure}

Fig.~\ref{fig:tmin_nstate_stability_2s} shows the dependence on the axial-vector matrix element fit parameter, $\mathring{g}_A$, under varying two-point and FH $\tsep^{\rm min}$, and their corresponding $n_{\textrm{states}}$. Typically, stability plots demonstrate that the best fit lies at a locally optimal point in the parameter space. The simplest strategy is to simply fix all but one parameter, such as $\tsep^{\rm min}$ or $n_{\textrm{states}}$ in this section. If results surrounding the best fit are absent of spurious correlations, then the stability plots provide evidence that systematic errors arising from the fit procedure are all accounted for. However, in the case of the FH form of the correlation function, the time dependence observed is analogous to a two-point correlation function. And similar to the two-point function, the best fit for a given number of excited states, requires a more careful choice of $\tsep^{\rm min}$. For example, a two-point fit with many excited states may successfully describe the data at small source-sink separation times, but will surely fail when the model is simplified. Therefore, for the stability plots shown in Fig.~\ref{fig:tmin_nstate_stability_2s}, the analysis varies $\tsep^{\rm min}$ around the best fit for a given number of states in the model. For simplicity, we vary the $\tsep^{\rm min}$ of the vector and axial-vector matrix elements in tandem.

After the summation, the three-point correlation function only has one degree of freedom per time slice. Coupled with the fact that the signal-to-noise is exponentially worse compared to the two-point function (see Fig.~\ref{fig:effective}), this means that in order to prevent over-fitting the data, simpler fit models with less excited states should be chosen. Therefore, the stability shown in Fig.~\ref{fig:tmin_nstate_stability_2s} is checked for fits with 1, 2, and 3 states. Beyond three states, the model will have comparable or more parameters than data. While this can in principle be alleviated with careful choices of priors, strategies that rely more heavily on prior information may also inadvertently introduce possible systematic errors.

For a single state, we observe the fit to be stable at approximately a source-sink separation time of 1 fm. This identifies the length scale where, given a percent-level determination of the matrix element, all excited states have decayed to below the noise. This observation is consistent with what is observed in our previous work~\cite{Bouchard:2016heu,Chang:2018uxx}. With two- and three-state fits, we observe that the best fit $\tsep^{\rm min}$ can encompass progressively shorter source-sink separation times, in agreement with expectation. Furthermore, the best fit for the three models, and neighboring results, are all consistent within one standard deviation with no appreciable systematic trend.

\begin{figure}[t]
\begin{center}
\includegraphics[width=\columnwidth]{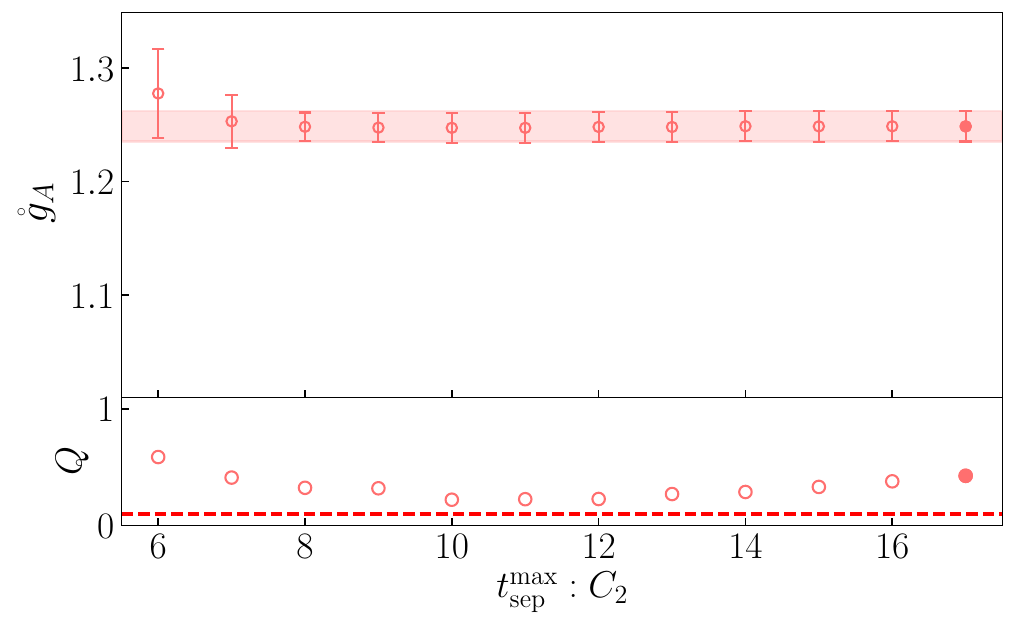}
\caption{\label{fig:tmax_nstate_stability_2s} Stability analysis for the axial coupling for the combined two-point and Feynman-Hellmann fit for varying $\tsep^{\rm max}$ of the two-point function. The best fit highlighted by a solid marker and corresponding horizontal band is identical to the one shown in Fig.~\ref{fig:tmin_nstate_stability_2s}. The corresponding $Q$-values are provided.}
\end{center}
\end{figure}

A two-state fit allows us to capture some of the excited state dependence in order to not rely entirely on the correlator reaching the plateau region. Simultaneously, fitting to two states avoids the possibility of over-fitting since an introduction of three additional fit parameters (ground state and first excited state matrix elements) can be extracted from 10 data points in this specific analysis ($\tsep^{\rm min}$ of 5 to 14). We assume that the overlap and energies are well constrained by the two-point correlation function in this counting. The three-state fit introduces 6 new matrix element parameters and is at the edge of what is naively allowed by data. Fitting to three states allows $\tsep^{\rm min}$ to encompass down to $\tsep=4$, resulting in 12 data points. However, since lattice correlation functions have an exponential signal-to-noise problem, data points do not carry equal weight in determining the posterior distributions. In particular, the weighted loss function penalizes larger time separation data with the inverse of the variance. Referring back to Fig.~\ref{fig:effective}, we see for this specific example, data beyond $\tsep=9$ have little impact on the outcome of the result.
This can be seen for example in Fig.~\ref{fig:tmax_nstate_stability_2s} where the results are given for varying $\tsep^{\rm max}$ of the two-point function.
As a result, the three-state fit with 6 extra parameters is effectively constrained by approximately 6 data points. It follows that more complicated fit functions run into the danger of over-fitting. We conclude that fits to the FH correlation function are best performed with two states.

\begin{center}{\textit{Two-point with three-point analysis} (\textbf{23})}\end{center}

\begin{figure}[t]
\begin{center}
\includegraphics[width=\columnwidth]{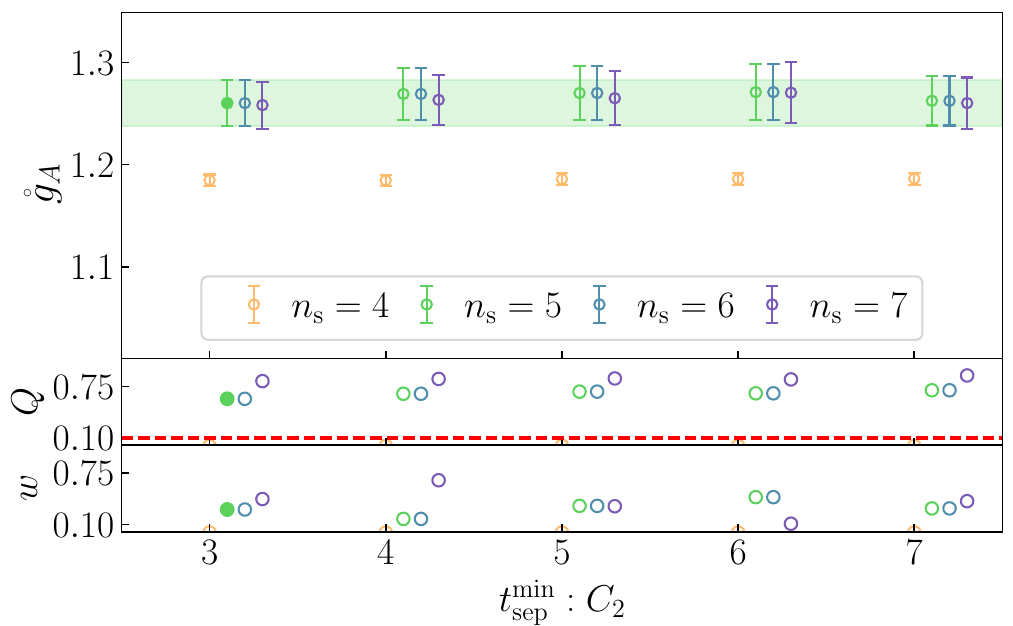}
\includegraphics[width=\columnwidth]{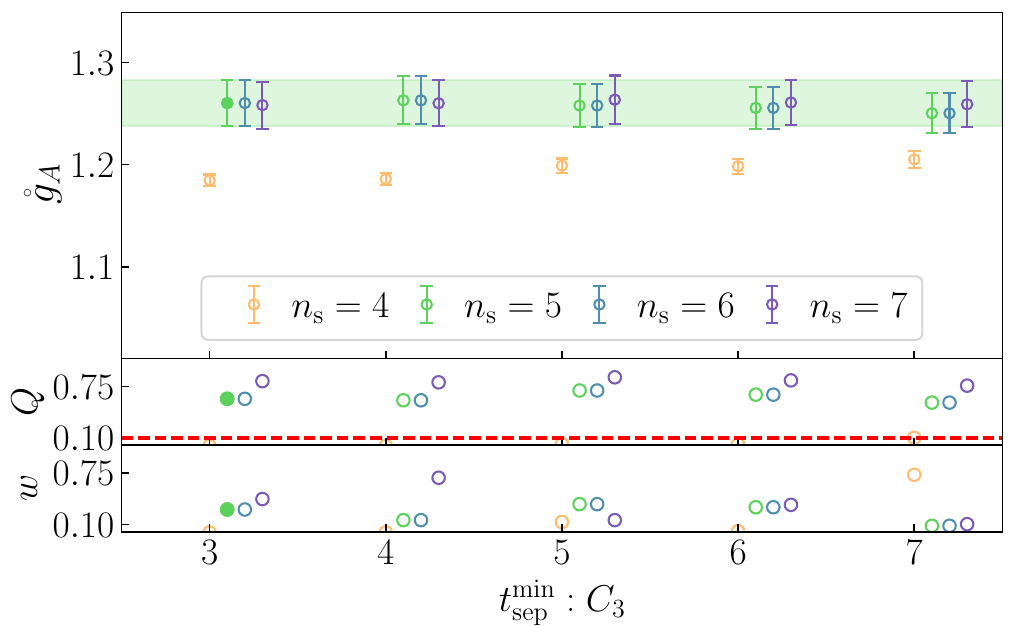}
\caption{\label{fig:tmin_nstate_stability_23} Stability analysis for $\mathring{g}_A$ for the combined two-point and three-point fit for (Top) varying $\tsep^{\rm min}$ of the two-point function and (Bottom) three-point correlation function. In both cases, results of $n_s=[4, 5, 6, 7]$ state fit ansatzes are shown. The filled marker and horizontal green band for the $n_s=5$ result highlights the best fit. The corresponding $Q$-values are provided. The Bayes Factors $w$ for a given $\tsep^{\rm min}$ are shown in the bottom panel of each plot.}
\end{center}
\end{figure}

\begin{figure}[t]
\begin{center}
\includegraphics[width=\columnwidth]{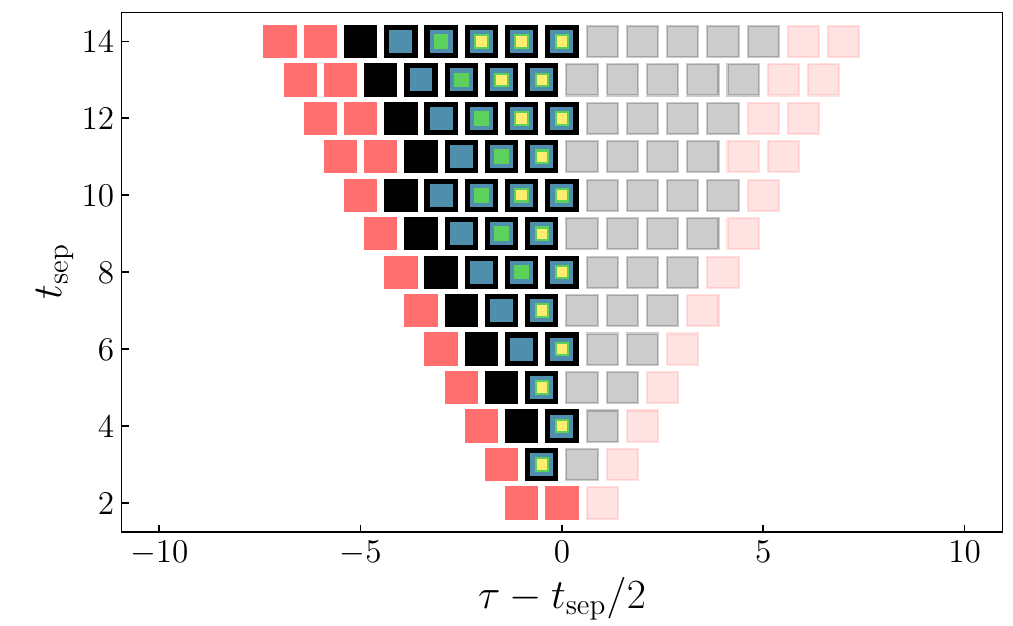}
\caption{\label{fig:3pt_fit_region}
This figure depicts the regions in $\tsep$ and $\t$ used in the \textbf{23} analysis.
The numerical results have been averaged about $\t=\tsep/2$ denoted by the light-shaded squares on the right half of the figure.
The outer-most (red) squares for each $\tsep$ are the points $\t=0$ or $\t=\tsep$ which are excluded from the analysis.
For $\tsep>10$, the values of $\t=1$ and $\t=\tsep-1$ are also excluded from the optimal fit.
Moving in, the (black) squares represent the values of $\t$ closest to the source/sink times included in the optimal analysis (with all ``inner'' values of $\t$ also included), $\t_{\rm inc}^{\rm opt}$.
Cutting an additional value of $\t$ closest to the source/sink times is $\t_{\rm inc}^{\rm opt} -1$ denoted by the lighter (blue) squares.
The lighter (green) squares denote $\t_{\rm inc}^{\rm opt} -2$ and the innermost (yellow) squares denote $\t_{\rm inc}^{\rm opt} -3$.
When cutting values of $\t$, we always keep one (for even $\tsep$) or two (for odd $\tsep$) values of $\t$ in the middle of the source-sink separation time.
}
\end{center}
\end{figure}

Rather than constructing the FH combination, we explore the possibility of fitting directly to the three-point correlator as a function of both source-sink and current-sink insertion time.
Fig.~\ref{fig:tmin_nstate_stability_23} shows the stability of the best fit for the two-point and three-point simultaneous fit under varying $\tsep^{\rm min}$ for the two-point and three-point function. Similar to the FH fit in the previous section, we simplify the analysis by varying the $\tsep^{\rm min}$ for the vector and axial-vector matrix elements simultaneously. The best fit presumably can be slightly improved if this condition is relaxed.
Due to the complexity involved in choosing a two-dimensional fit region, we further simplify the decision making process by fitting to all valid insertion-sink time separation times (the contact terms are dropped) as shown in Fig.~\ref{fig:3pt_fit_region}. This choice reduces the $\tsep^{\rm min}$ stability study to again a single dimension.

To capture this curvature, a large number of excited states are required.
This is corroborated by our analysis of the two-point correlation functions:
since we have used conservative smearing on the quark interpolating operators, fits to small source-sink separation times also require a large number of excited states.
Due to the large number of states, stability with respect to $\tsep^{\rm min}$ are less sensitive to $n_{\textrm{states}}$. As a result, we further simplify the study by showing only stability with respect to the final result, instead of first identifying $n_{\textrm{states}}$-dependent best fits shown in Fig.~\ref{fig:tmin_nstate_stability_2s}.

The best fit, which includes nearly all current insertion times (aside from the source and sink time) is observed to require a 5 state model and includes source-sink separation times that are commensurate to the inclusion of small current-sink times.
Unlike the FH correlator, the three-point correlator supplies many data points as shown in Fig.~\ref{fig:psychedelic_moose}, and can be used to fully capture the complicated excited-state structure.
The best fit lies in the region of stability, and is chosen to incorporate the most data possible given a fit model. At the same time, for a given fit region, the simplest model is chosen (fewest number of states). For example, the best fit with a three-point $\tsep^{\rm min}=3$ includes 5 states even though 6 or 7 states yields similar ground state posterior distributions. This decision is corroborated with the set of Bayes Factors normalized to a fixed fit region, \eqnref{eq:w-logGBF}.  For example, Fig.~\ref{fig:tmin_nstate_stability_23} suggests that 5 state and 7 state fit have comparable probability for reproducing the underlying data.

\begin{figure}[t]
\begin{center}
\includegraphics[width=\columnwidth]{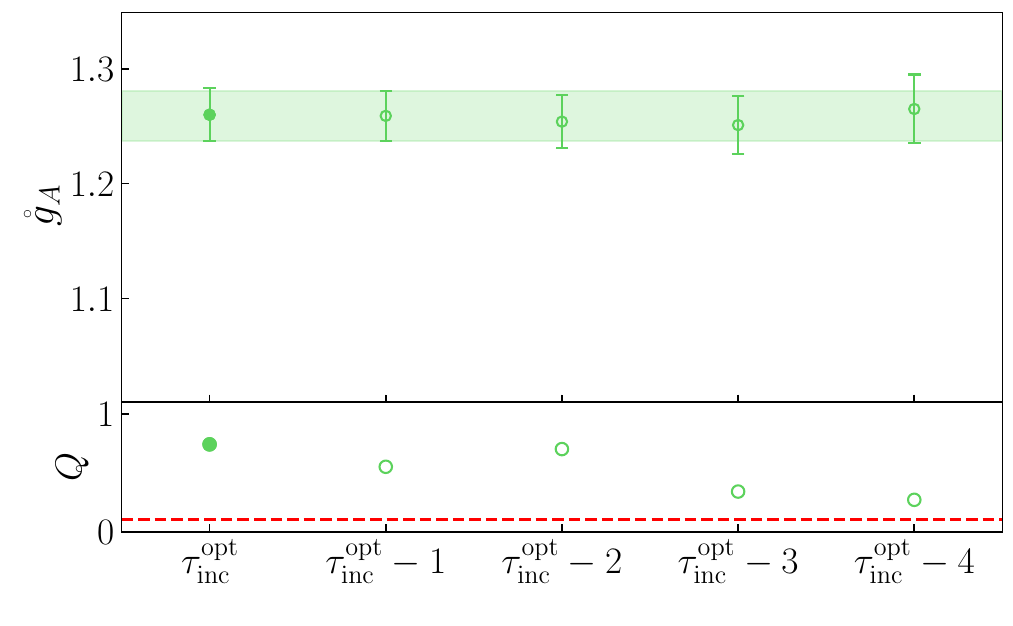}
\caption{\label{fig:tins_stability_23} Stability analysis for the combined $C_2(\tsep)$ and $R_\G(\tsep,\t)$ analysis for varying regions of current-insertion time. The best fit given by $\tau^{\textrm{opt}}_{\textrm{inc}}$ is defined in Table~\ref{tab:tsep_range} (see also \figref{fig:3pt_fit_region} for a graphical depiction of $\tau^{\textrm{opt}}_{\textrm{inc}}-n$).
The filled marker and horizontal green band highlights the best fit and is identical to the best fit in Fig.~\ref{fig:tmin_nstate_stability_23}. The $Q$-values are provided.}
\end{center}
\end{figure}

While we make the choice of fitting data over all current insertion times, the ground-state posterior distributions are stable when extracted from a subset of the data. In Fig.~\ref{fig:tins_stability_23}, the fit region with respect to current insertion time is symmetrically truncated, keeping however, at least one data point per source-sink separation time (\textit{e.g.} $\tsep=3$ includes $\t=[1,2]$, so a $\t_c=1$ will not eliminate any data since otherwise the entire dataset for $t=3$ would be eliminated). It is also observed that while curvature in $\tau$ is dependent predominantly on excited-state behavior, aggressively truncating data still leads to larger statistical uncertainty in ground state parameters since less data is being included in the analysis.

We conclude that a simultaneous fit to the two- and three-point correlators is best performed by fitting to the maximum amount of data while choosing the simplest model which can describe the data. In particular, we observe that the under this strategy, the accompany two-point correlator also provides sufficient constraints on the excited state overlap and energy parameters.

\begin{center}{\textit{Two-point, three-point and FH Analysis} (\textbf{23s})}\end{center}

\begin{figure}[t]
\begin{center}
\includegraphics[width=\columnwidth]{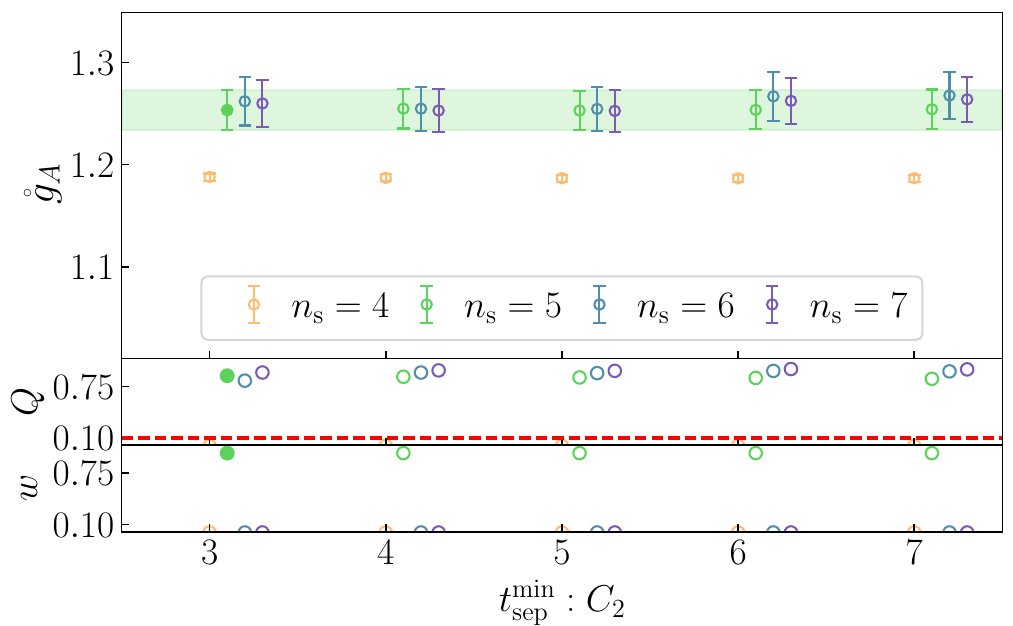}
\includegraphics[width=\columnwidth]{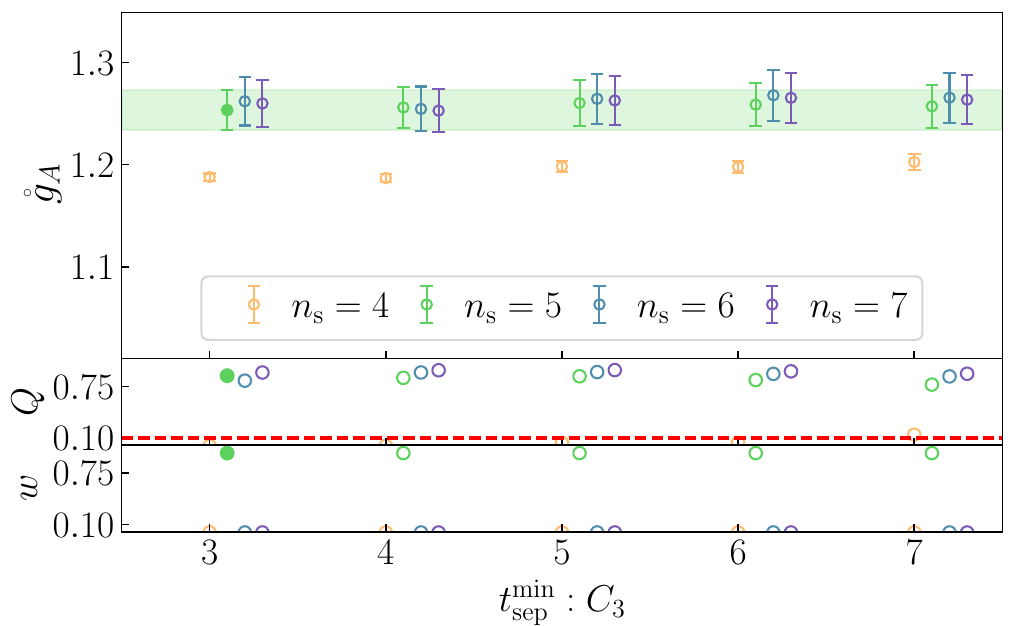}
\includegraphics[width=\columnwidth]{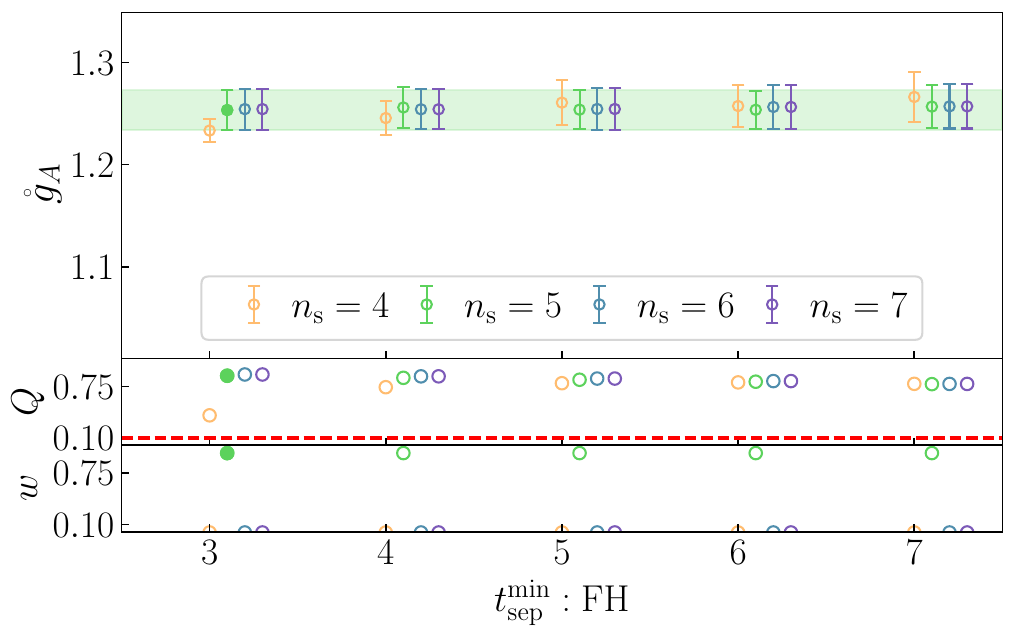}
\caption{\label{fig:tmin_nstate_stability_23s} Stability analysis for the combined two-point, three-point, and Feynman-Hellmann analysis. These plots follow what is shown in Fig.~\ref{fig:tmin_nstate_stability_23}.}
\end{center}
\end{figure}

Finally, we perform a simultaneous fit to the two-point correlator along with all FH and three-point correlators. Since the FH correlator exposes different excited-state dependence when compared to the three-point correlator, it may be possible to extract a more robust calculation of the ground state parameters. Unfortunately, the overall strategy of the previous two studies are incompatible with one another. In the case of the FH analysis, the strategy is to fit simpler models in order to avoid overfitting, while for the three-point analysis overfitting is much less of a concern and instead a majority of clean data are fit with more complex models. A successful combined fit will need to reconcile these differences.

We take a simple approach by recognizing that the excited state information extracted from the three-point analysis can be used to constrain a more complex FH fit function. Following this logic, the best combined fit follows the same 5 state model as the two-point with three-point correlator fit discussed previously, while the FH fit is now modelled by 5 states with the intention of relying on the three-point correlator to constrain high excited state parameters. Fig.~\ref{fig:tmin_nstate_stability_23s} demonstrates the stability of the combined best fit under changes in $\tsep^{\rm min}$ of the two-point, three-point and FH correlators. We observe that in the combined fit the ground-state parameters are insensitive to changes in $\tsep^{\rm min}$ for all datasets, including the FH correlation functions. This observation corroborates the hypothesis that the three-point correlator lends support to high excited-state contributions which is consistent with the predicted spectral decomposition for both correlation functions. Additionally, while the best fit $\tsep^{\rm min}$ for the three-point is kept the same as the three-point with two-point fit, the best-fit FH $\tsep^{\rm min}$ now extends down to $\tsep=3$ (previously, in the \textbf{2s} best fit $\tsep^{\rm min}$ is 5) due to the inclusion of more states in the model.

\begin{figure}[t]
\begin{center}
\includegraphics[width=\columnwidth]{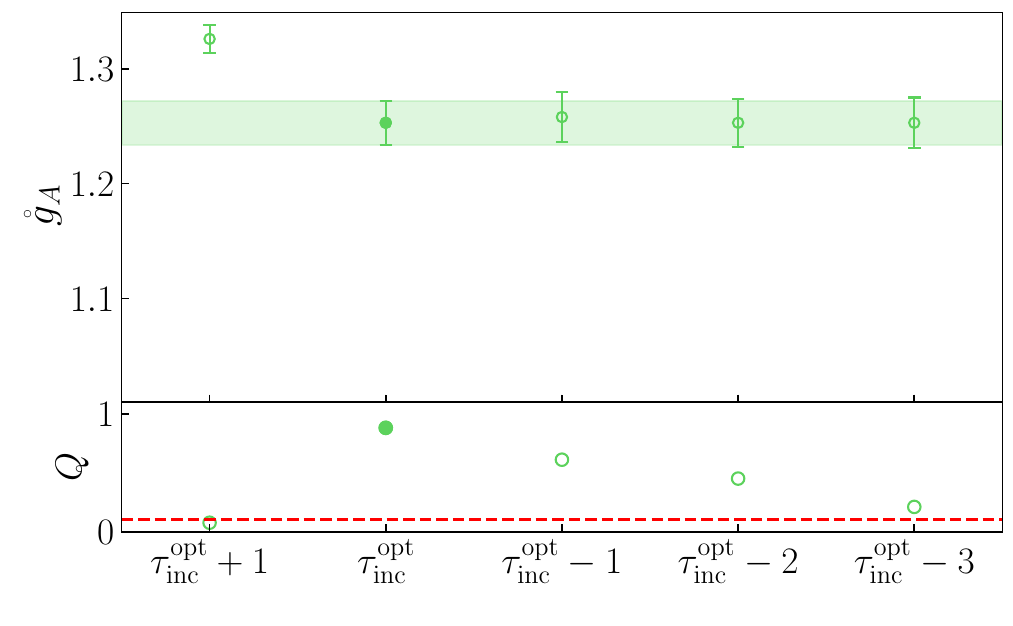}
\caption{\label{fig:tins_stability_23s} Stability analysis for the combined two-point, three-point, and Feynman-Hellmann analysis for varying regions of current-insertion time. This plot follows what is shown in Fig.~\ref{fig:tins_stability_23}.}
\end{center}
\end{figure}

Finally, sensitivity under varying current insertion time for the three-point correlator is studied. Unlike the simpler three-point with two-point correlator analysis, we have to drop one additional data point away from the contact interaction at $t=11$ to $t=14$ for the ground-state posterior distributions to be insensitive to changes in fit region. Fig.~\ref{fig:tins_stability_23s} shows the varying fit region with respect to the best fit. In particular, the $\t^{\rm opt}_{\rm inc} +1$ fit is to all current insertion dependence between the source and sink, $\t=[1,\tsep-1]$ for all $\tsep$. The colored regions in Fig.~\ref{fig:3pt_fit_region} highlight the various $\t_{\rm inc}$ regions with the colored boxes. The black region denotes the best fit region. We observe that the best fit lies in a region that is insensitive to varying subsets of the three-point correlator. Conversely, the combined fit suggests that at large values of $\tsep$, where the distribution of the correlator become under-sampled, the three-point data shows signs of being inconsistent with the FH data, leading to the instability seen in $\t^{\rm opt}_{\rm inc} +1$ fit.

We conclude that the three-point correlator provides enough information to constrain the excited states of the FH correlator. A combined fit is therefore, best performed by rooting the calculation in a two-point with three-point strategy and extending the analysis to encompass as much of the FH correlator as is describable by the truncated spectral decomposition.

\subsubsection{Prior width analysis}

\begin{figure}[t]
\begin{center}
\includegraphics[width=\columnwidth]{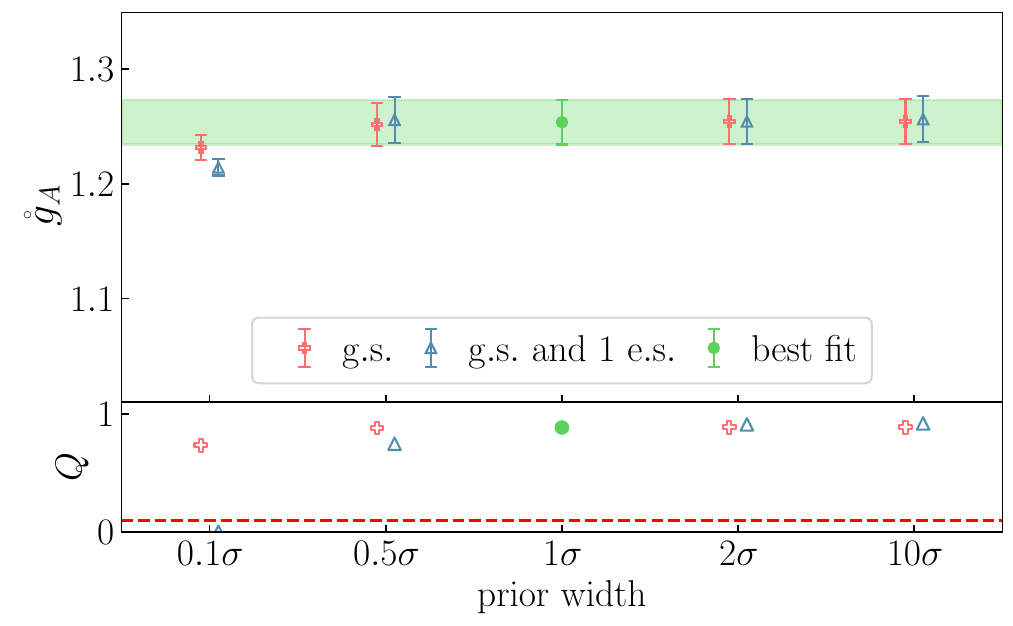}
\caption{\label{fig:prior_width} Sensitivity analysis of the best fit result from the combined two-point, three-point, and Feynman-Hellmann fit under varying prior widths for the ground-state (red) and together with the first-excited state (blue). The $x$-axis indicates that the study performed variations from 0.1 to 10 times the best fit prior width. The filled marker and horizontal green band highlights the best fit and is identical to the result shown in Fig.~\ref{fig:tmin_nstate_stability_23s}.}
\end{center}
\end{figure}

Our overall strategy is to extract ground-state parameters from a multi-exponential fit, subject to prior constraints. Since the objective function now depends on prior distributions in addition to data, we check that the posterior distributions of interest are insensitive to our prior knowledge. The purpose of introducing priors in this context is not to supplement additional information, but to constrain the search space of the numerical minimization for faster convergence. Lower computation costs allow us to more thoroughly investigate the sensitivity of ground state posterior distributions, which in turn lends to more robust results. We enforce the expectation that choices of priors should not yield changes in the posterior distributions of ground-state parameters.

Fig.~\ref{fig:prior_width} demonstrates the robustness of our quoted results under variations of the ground-state and first-excited state prior widths for the combined fit. The study indicates that the extracted matrix elements are unconstrained by prior distributions until the widths are reduced by a factor of 10, while broadening the prior distribution by a factor of 10 leaves the matrix elements unchanged. Similar conclusions hold for the vector matrix element for the \textbf{2s}, and \textbf{23} strategies.

\subsubsection{Consistency of spectrum and matrix elements from different correlation functions\label{app:spec_g_consistency}}
\begin{figure*}
\begin{center}
\includegraphics[width=\textwidth]{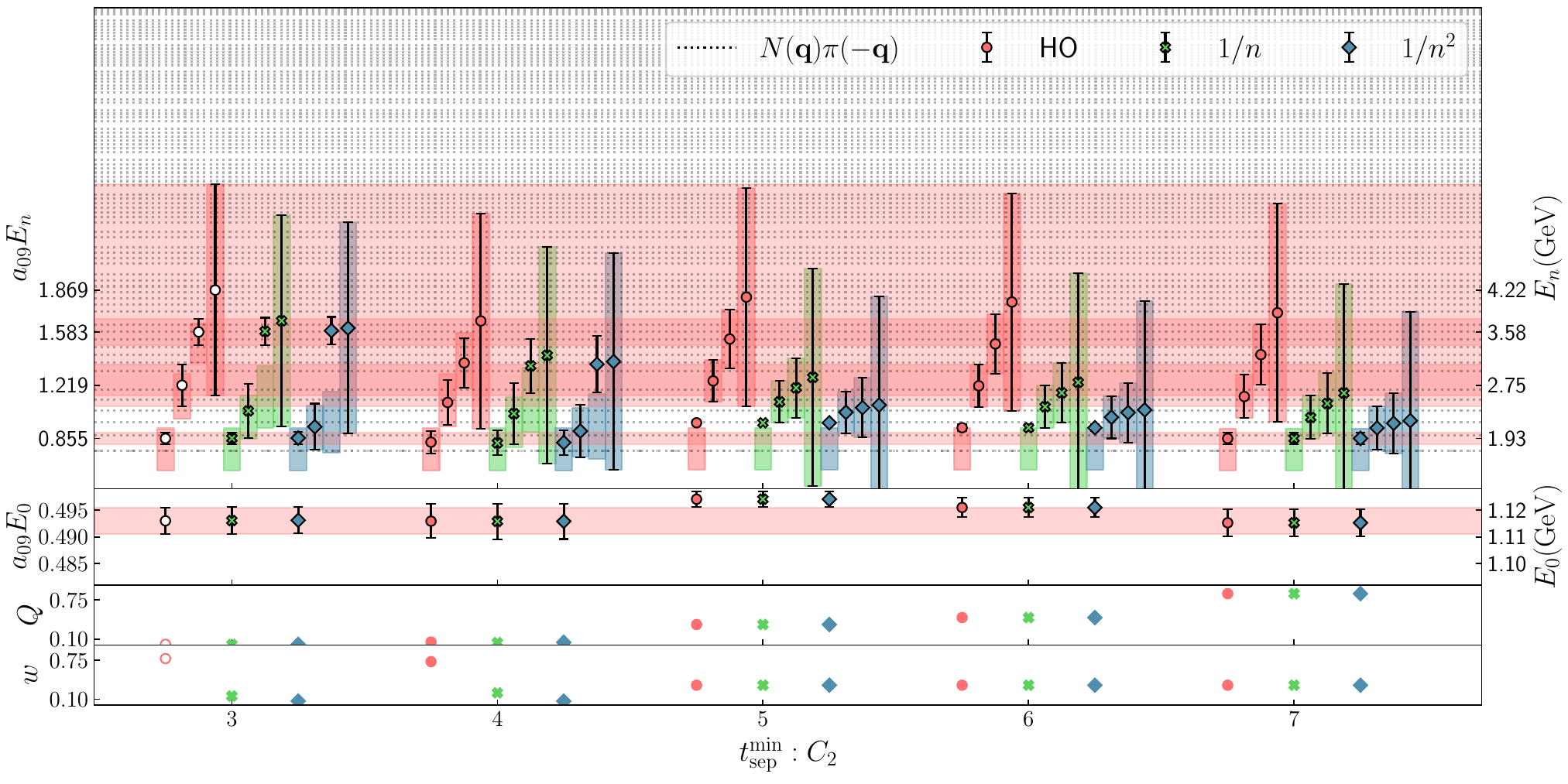}
\includegraphics[width=\textwidth]{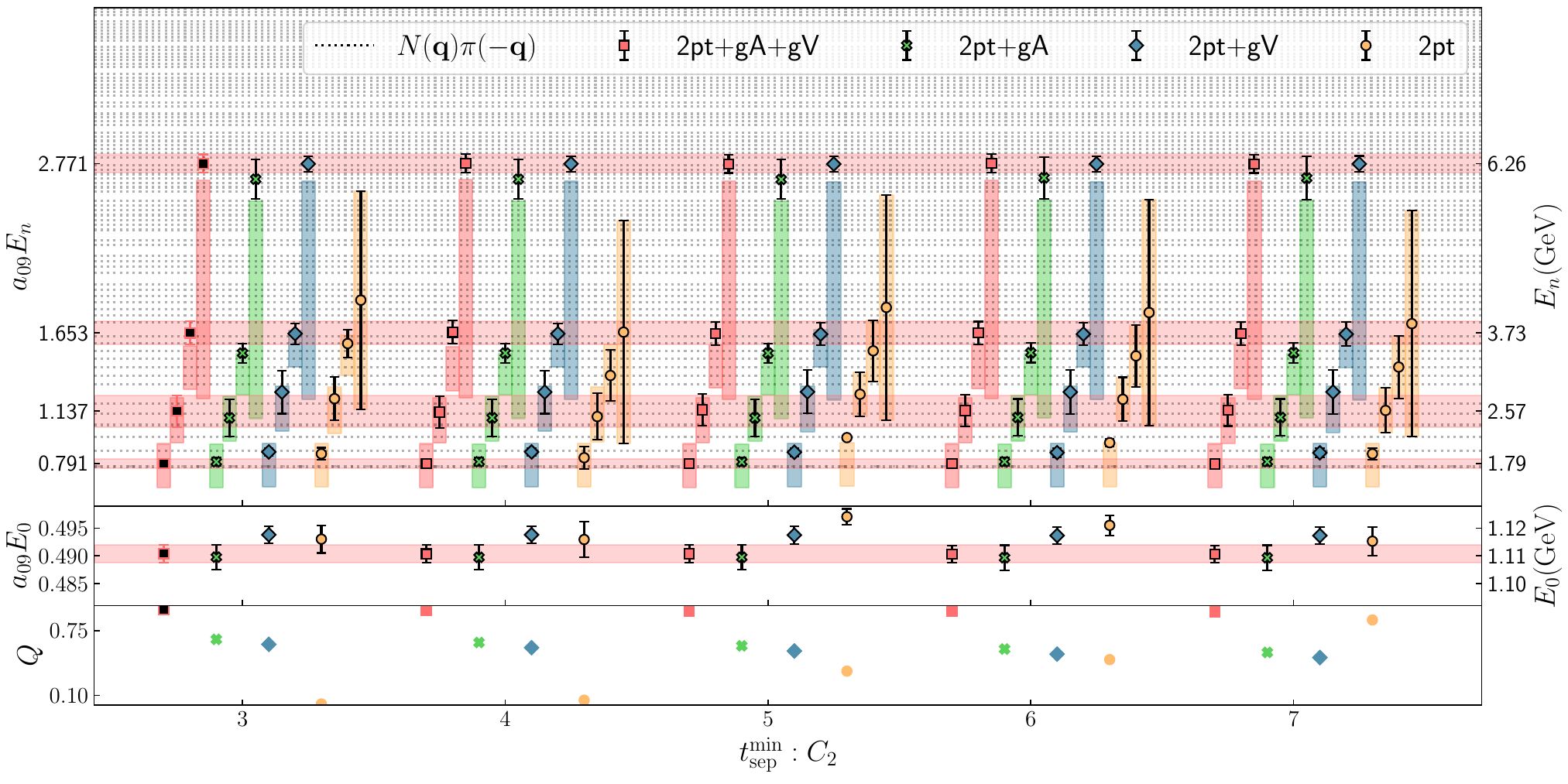}
\caption{\label{fig:spec_g_consistency} Consistency of the extracted spectrum and matrix elements from various analyses.
In the top figure, we show the extracted spectrum as a function of the excited state model and $\tsep^{\rm min}$ of the two-point function.  The horizontal (red) bands are from the HO analysis with $\tsep^{\rm min}=3$, to guide the eye.
In the bottom figure, we show the HO model of excited states for various global minimizations.  From left to right (for each $\tsep^{\rm min}$), we show the full analysis (2pt+gA+gV), a fit to the two-point functions (2pt), a fit to the two-point and $A_3$ correlation function (2pt + gA) and a fit to the two-point and $V_4$ correlation function (2pt+gV).}
\end{center}
\end{figure*}

In \figref{fig:spec_g_consistency}, we show the consistency (or lack thereof) of the excited and ground state spectrum from various global minimzations of different sets of correlation functions.
One observes that the analysis of the two-point correlation function is not sufficient to constrain the excited state spectrum: for most results, the posterior energy levels simply follow their priors.
In contrast, when either the axial or vector three-point function (or both) are simultaneously analyzed (with the full data-covariance), the excited state spectrum is much more precisely constrained, often in tension with the priors.  Further, once a global minimization is performed, the values of the excited state spectrum become significantly more consistent.

\section{Prior and posterior distributions\label{app:priors}}
In \tabref{tab:2pt_priors}, we list the prior and posterior distributions of the energies and overlap factors used in our final analysis.
The matrix element prior and posterior distributions are given in \tabref{tab:g_priors}.
We use the $(\mu,\s)$ format to report the central value $\mu$ and the width $\s$ of the distributions.
In \tabref{tab:tsep_range}, we list the range of $\tsep$ values used in the three sets of correlation functions that are analyzed.

\begin{table}
\caption{\label{tab:2pt_priors}
We list the prior and posterior energies and overlap factors, in lattice units, for each of the parameters that describe the two-point correlation function.  The prior and posterior values are listed in $(\mu,\s)$ pairs.
The energy splittings are priored with a log-normal distribution with $\D E_n = (2m_\pi, m_\pi)$, except for the highest state which has a prior width of $5m_\pi$, with the total energy given by the sum in \eqnref{eq:spectrum_parameterization}.  For the optimal fit, all energy splittings are given a central value of $2m_\pi$.
}
\begin{ruledtabular}
\begin{tabular}{r c c}
Parameter& Prior & Posterior \\
\hline
$E_0$  & $(0.50, 0.05)$ & $(0.4904, 0.0016)$ \\
$\D E_{n<4}$ & $(0.29, 0.14)$ & $(0.3, 0.028)$ \\
$E_1$  & $(0.79, 0.15)$ & $(0.79, 0.028)$ \\
$E_2$ & $(1.07, 0.21)$ & $(1.136, 0.099)$ \\
$E_3$ & $(1.36, 0.25)$ & $(1.66, 0.16)$ \\
$\D E_4$     & $(0.29, 0.72)$ & $(1.119, 0.033)$\\
$E_4$ & $(1.65, 0.76)$ & $(2.78, 0.16)$ \\
$z_0$ & $(0.00034, 0.00034)$ & $(0.0003211, 0.0000039)$ \\
$z_1$ & $(0, 0.00025)$ & $(0.000333, 0.000031) $ \\
$z_2$ & $(0, 0.00015)$ & $(-0.000331, 0.000062)$ \\
$z_3$ & $(0, 0.00015)$ & $(0.000617, 0.000034)$ \\
$z_4$ & $(0, 0.00015)$ & $(0.00032, 0.00011)$ \\
\end{tabular}
\end{ruledtabular}
\end{table}

\begingroup \squeezetable
\begin{table}
\caption{\label{tab:g_priors}
Priors and posteriors of the vector and axial-vector matrix elements.
The values are listed in $(\mu,\s)$ pairs.
}
\begin{ruledtabular}
\begin{tabular}{ccc|ccc}
$g^A_{nm}$ & Prior & Posterior &$g^V_{nm}$ & Prior & Posterior \\
\hline
$g^A_{00}$ & $(1.2, 0.2)$ & $(1.253, 0.019)$ & $g^V_{00}$ & $(1, 0.2)$ & $(1.02238, 0.00087)$ \\
$g^A_{01}$ & $(0, 1)$ & $(-0.271, 0.068)$    & $g^V_{01}$ & $(0, 1)$ & $(-0.0041, 0.004)$     \\
$g^A_{11}$ & $(0, 1)$ & $(0.89, 0.2)$        & $g^V_{11}$ & $(1, 0.2)$ & $(0.997, 0.02)$      \\
$g^A_{02}$ & $(0, 1)$ & $(-0.16, 0.11)$      & $g^V_{02}$ & $(0, 1)$ & $(0.002, 0.013)$       \\
$g^A_{12}$ & $(0, 1)$ & $(-0.05, 0.3)$       & $g^V_{12}$ & $(0, 1)$ & $(0.013, 0.042)$       \\
$g^A_{22}$ & $(0, 1)$ & $(0.71, 0.53)$       & $g^V_{22}$ & $(1, 0.2)$ & $(1.05, 0.12)$       \\
$g^A_{03}$ & $(0, 1)$ & $(0.1, 0.043)$       & $g^V_{03}$ & $(0, 1)$ & $(0.0798, 0.0075)$     \\
$g^A_{13}$ & $(0, 1)$ & $(0.14, 0.14)$       & $g^V_{13}$ & $(0, 1)$ & $(0.071, 0.025)$       \\
$g^A_{23}$ & $(0, 1)$ & $(0.25, 0.31)$       & $g^V_{23}$ & $(0, 1)$ & $(-0.011, 0.062)$      \\
$g^A_{33}$ & $(0, 1)$ & $(0.87, 0.22)$       & $g^V_{33}$ & $(1, 0.2)$ & $(0.789, 0.047)$     \\
$g^A_{04}$ & $(0, 1)$ & $(0.61, 0.21)$       & $g^V_{04}$ & $(0, 1)$ & $(0.35, 0.12)$         \\
$g^A_{14}$ & $(0, 1)$ & $(0.4, 0.25)$        & $g^V_{14}$ & $(0, 1)$ & $(0.33, 0.13)$         \\
$g^A_{24}$ & $(0, 1)$ & $(-1.06, 0.52)$      & $g^V_{24}$ & $(0, 1)$ & $(-0.34, 0.16)$        \\
$g^A_{34}$ & $(0, 1)$ & $(0.72, 0.39)$       & $g^V_{34}$ & $(0, 1)$ & $(0.63, 0.24)$         \\
$g^A_{44}$ & $(0, 1)$ & $(-0.32, 0.95)$      & $g^V_{44}$ & $(1, 0.2)$ & $(1, 0.2)$           \\
\end{tabular}
\end{ruledtabular}
\end{table}
\endgroup

\begin{table}
\caption{\label{tab:tsep_range}
We list the range of $\tsep$ values used in our optimal fits for each of the three sets of correlation functions we analyze.  We also list the values of the current insertion time, $\t$, for which the optimal fit uses $\t=[1,\tsep-1]$ for the \textbf{23} analysis (2pt and 3pt).  When we perform the \textbf{23s} (2pt, 3pt and FH) analysis, there is mild tension when including the current times $\t=1$ and $\t=\tsep-1$ for $\tsep>10$ and so our optimal fit removes one extra current insertion time for these later $\tsep$ values.
For FH, the $\t$ range indicates the values used in the summation over current time.
}
\begin{ruledtabular}
\begin{tabular}{r c c c}
& (2pt, 3pt) & (2pt, FH) & (2pt, 3pt, FH)\\
\hline
2pt $\tsep$ range & [3, 17] & [5, 17] & [3, 17] \\
3pt $\tsep$ range & [3, 14] & --      & [3, 14] \\
FH  $\tsep$ range & --      & [5, 13] & [3, 13] \\
3pt $\t$    range & [1,$\tsep-1$]& [1,$\tsep-1$] & [1,$\tsep-1$], $\tsep\leq10$\\
                  &              & [1,$\tsep-1$] & [2,$\tsep-2$], $\tsep>10$\\
FH  $\t$    range & --      & [1,$\tsep-1$]& [1,$\tsep-1$]\\
$n_{\textrm{states}}$        & 5       & 2       & 5
\end{tabular}
\end{ruledtabular}
\end{table}

\bibliography{ga_excited_states}

\end{document}